\documentclass[]{tPHM2e}

\usepackage{graphicx}
\usepackage[usenames,dvipsnames]{xcolor}
\usepackage{epsfig}
\usepackage{epstopdf}

\newcommand{\avg}[1]{\langle #1 \rangle}

\begin{document}

\markboth{C. B. O'Donovan et al.}{}

\title{Mean-field granocentric approach in 2D \& 3D polydisperse, frictionless packings}

\author{C. B. O'Donovan$^{\rm a}$$^{\ast}$\thanks{$^\ast$Corresponding author. Email: caodonov@tcd.ie
\vspace{6pt}}, E. I. Corwin$^{\rm b}$\vspace{6pt} and M. E. M\"{o}bius$^{\rm a}$\\
 $^{\rm a}${\em{School of Physics, Trinity College Dublin, Dublin 2, Ireland}};
\newline$^{\rm b}${\em{Department of Physics, University of Oregon, Eugene, Oregon 97403, USA}}\\\vspace{6pt}\received{Version submitted: 8$^{th}$February 2013} }

\maketitle

\begin{abstract}
We have studied the contact network properties of two and three dimensional polydisperse, frictionless sphere packings at the random closed packing density through simulations. We observe universal correlations between particle size and contact number that are independent of the polydispersity of the packing. This allows us to formulate a mean field version of the granocentric model to predict the contact number distribution $P(z)$. We find the predictions to be in good agreement with a wide range of discrete and continuous size distributions. The values of the two parameters that appear in the model are also independent of the polydispersity of the packing. Finally we look at the nearest neighbour spatial correlations to investigate the validity of the granocentric approach. We find that both particle size and contact number are anti-correlated which contrasts with the assumptions of the granocentric model. Despite this shortcoming, the correlations are sufficiently weak which explains the good approximation of $P(z)$ obtained from the model.
\bigskip

\begin{keywords}Packings, Disorder, Spatial correlations
\end{keywords}\bigskip

\end{abstract}

\section{Introduction}
The question of how spheres pack together has been of interest to scientists for centuries\cite{PerfectPacking}. In the context of amorphous materials the jamming transition of random close packed spheres are of particular interest and have been a substantial area of study in recent years\cite{Epitome,PackRevVanHecke,Anderson,Torquato,WyartRev}. At the jamming point, which corresponds to a critical packing density, the packing makes a sharp transition towards a mechanical stable state. At this point the isostatic condition requires that the average coordination number of the packing is twice the number of dimensions of the system \cite{Alexander}. This density is referred to as the random close packing density $\phi_{RCP}$.

\begin{figure}
\begin{center}
\includegraphics[width=120.0mm]{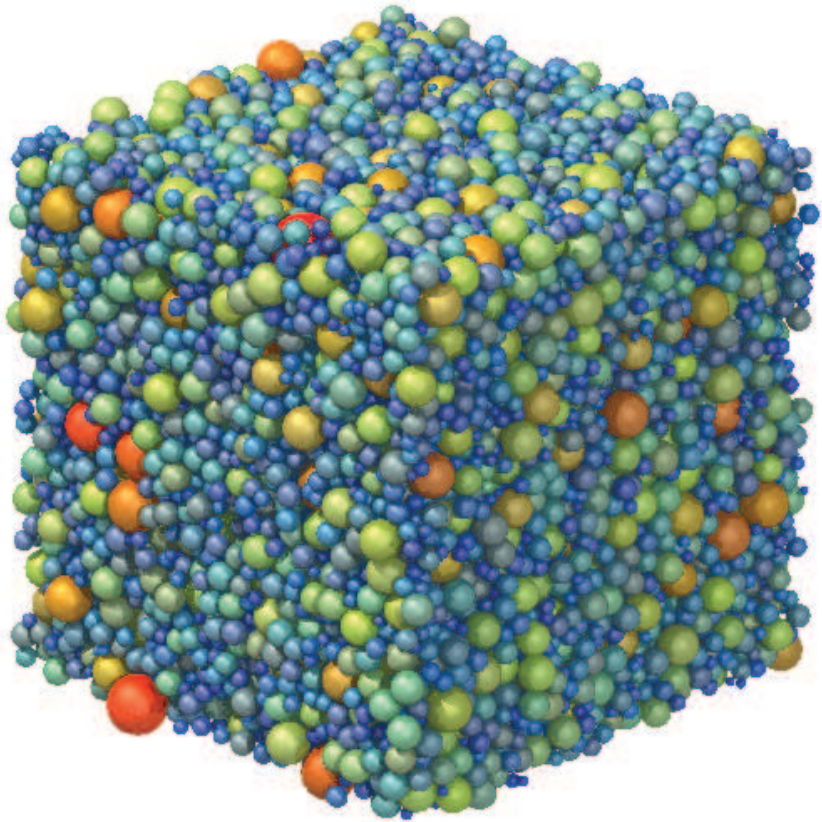}
\caption{\label{figpic}Visualisation of soft sphere packing simulation at $\phi_{RCP}$ with a lognormal distribution of radii. The spheres are coloured with a spectrum going from blue to red according to size with blue correspond to the smallest particles.}
\end{center}
\end{figure}

The packing of equal sized spheres in disordered configurations have a long history\cite{Bernal} and the value at which $\phi_{RCP}$ is reached for these packing is well studied in both experiment \cite{Bernal, Scott} and simulations\cite{Berryman, Epitome, SongMakse}, though $\phi_{RCP}$ has been shown to be dependent upon the history of the packing and the packing protocol used\cite{CorwinMicrostructure,Torquato}. More prevalent in nature, though not as a widely studied, are packings with a distribution of sizes. Experiments and simulations of binary mixtures\cite{Yerazunis,Kansal,ClarkeWiley,OHernRCP} and continuous size distributions\cite{Schaertl,HePolySim,GrootRCP,ChaikinPoly,GranocentricN,GranocentricSM} have investigated the value of $\phi_{RCP}$ and found that it increases with polydispersity. There have been some simulation and experimental studies on the contact properties of polydisperse, disordered packings\cite{HePolySim,BrujicForceDist,KatgertJamGeom} and recently the granocentric model has been proposed to predict the local packing structure at $\phi_{RCP}$\cite{GranocentricN,GranocentricSM,GranoMono} in three dimensions.

In this report we investigate the correlations between size and contact number of particles in polydisperse packings at $\phi_{RCP}$ in two and three dimensions for a wide range of size distributions. Our key finding is the existence of universal correlations between size and contact number that is \emph{independent} of the polydispersity. This empirical result allows us to formulate a mean field approach based on the granocentric model that yields excellent agreement with our data. One of the key assumption in the granocentric model is the lack of spatial correlations of both size and contact number of the particles. Our measurements of nearest neighbour correlations show that this assumption is violated. In general, the average contact number and the average size of neighbouring particles do not correspond to the global mean of contact number and size. In 3D packings larger particles are surrounded by smaller particles and vice versa. Moreover, particles with few contacts are neighbouring particles with many contacts. Nevertheless, these correlations are weak enough so that the predictions we obtain from the granocentric model agree well with our data.

\section{Simulations}

We model the disordered packings at $\phi_{RCP}$ through simulation of soft spheres. These are frictionless spheres that interact through purely repulsive body centred forces, which can be written as a function of the overlap between two particles in contact. The overlap is
\begin{equation}
\delta_{ij} = 1- \frac{d_{ij}}{R_i + R_j},
\end{equation}
where $R_{i}$ and $R_{j}$ are the radii of spheres $i$ and $j$ and $d_{ij}$ is the distance between the respective centres of the spheres.
The  interaction potential of the spheres is
\begin{equation}
V ( d_{ij} ) =\begin{cases}\frac{k}{2}\delta_{ij}^{2}, & \mbox{if }  \delta_{ij} > 0,\\
0, & \mbox{otherwise.}\end{cases}
\end{equation}
These interactions are harmonic with a spring constant $k$.
The spheres have their radius drawn from a set size distribution and are placed at random in a three dimensional periodic cell. The radii of the spheres are then rescaled such that the desired packing fraction $\phi$ is reached. A conjugate gradient method is then used to minimise the overlap between spheres and hence the the total energy of the packing\cite{NumericalRecipes}. The simulation is halted when the packing is in a local energy minimum and in mechanical equilibrium.

\begin{figure}
\begin{center}
\includegraphics[angle=270,width=120.0mm]{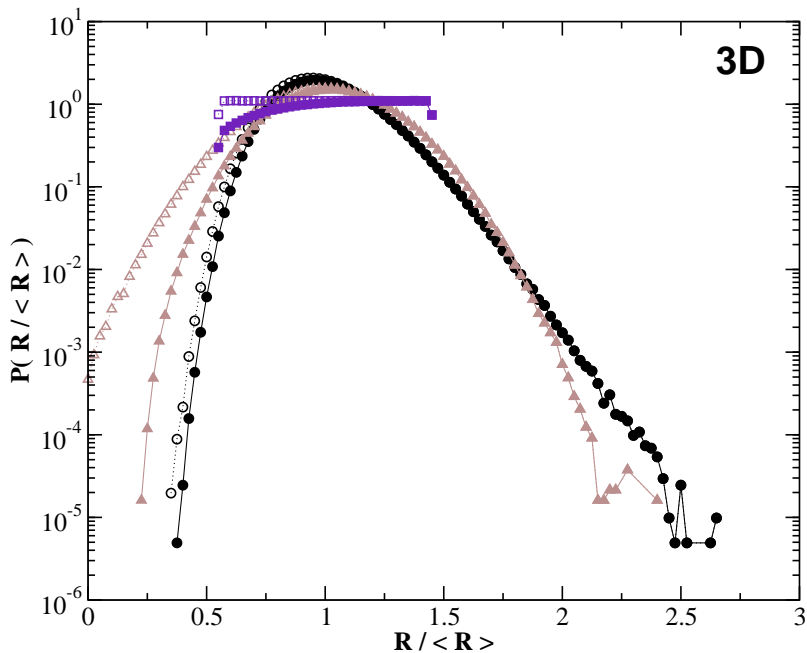}
\caption{\label{figrdist} Continuous size distributions used to create the soft sphere packings. The distributions are: (\textcolor{black}{$\bigcirc$}) lognormal $\sigma_A = 0.40$; (\textcolor{brown}{$\vartriangle$}) Gaussian $\sigma_A = 0.44$; (\textcolor{Purple}{$\square$}) uniform $\sigma_A = 0.44$. The open symbols represent the original size distribution and the closed symbols represent the size distribution once rattlers are removed.}
\end{center}
\end{figure}

In general, the isostatic condition can be shown by the following argument. If there are $D$ dimensions with $N$ soft particles, at $\phi_{RCP}$ there will be an average number of contacts $\avg{z}$ (where $\avg{.}$ denotes an average over all particles). Therefore in total there will be $N\avg{z} / 2$ contacts in the packing since every contact is shared by two particles. For mechanical stability all the contact forces need to balance on each particle\cite{Alexander}, which leads to matching the $ND$ degrees of freedom with the the contact forces giving
 \begin{equation}
 \label{eqIso}
ND = \frac{N  \avg{z}}{2}.
\end{equation}
The isostatic point $z_c$, equivalent to $\phi_{RCP}$, which is defined as when
\begin{equation}
z_{c} =\begin{cases}6, & \mbox{in 3 dimensions},\\
4, & \mbox{in 2 dimensions.}\end{cases}
\end{equation}
from Equation (\ref{eqIso}). While globally these mechanically jammed states are constrained to have $\avg{z} = z_c$,  there is a distribution of contact numbers for particles.

In general particles that have less than $D+1$ contacts cannot be locally mechanical stable. For 3D that means all particles with less than $4$ contacts and 2D, all particles with less than $3$ contacts are locally unstable. These particles are called rattlers and their contribution to the contact number analysis is omitted as their contact number is ill-defined. In a recent publication \cite{vanHeckeShearUnstable} it was shown that packings which satisfy Equation (\ref{eqIso}) can still be unstable under shear, though this effect is only pronounced for systems with a low number of particles in the packing. This effect is negligible in our simulations.

Each of the simulated packings has $16384$ particles with various different size distributions, with up to $500$ realisations in three dimensions and $50$ realisations in two dimensions for each size distribution. An example of a sphere packing is shown in Figure \ref{figpic}. A variety of size distributions are created including discrete size distributions of monodisperse and bidisperse spheres, where there is a 50-50 mixture with a size ratio 1:1.4, and continuous radius distributions such as the lognormal distribution, Gaussian distribution and uniform distribution, which are plotted in Figure \ref{figrdist}. Packings at $\phi_{RCP}$ are found by starting with a packing density above $\phi_{RCP}$ which is lowered until the average contact number $\avg{z}$ is within the range $z_c + 0.05 > \avg{z} \geq z_c$.

\begin{figure}
\centering
\includegraphics[angle=270,width=68.0mm]{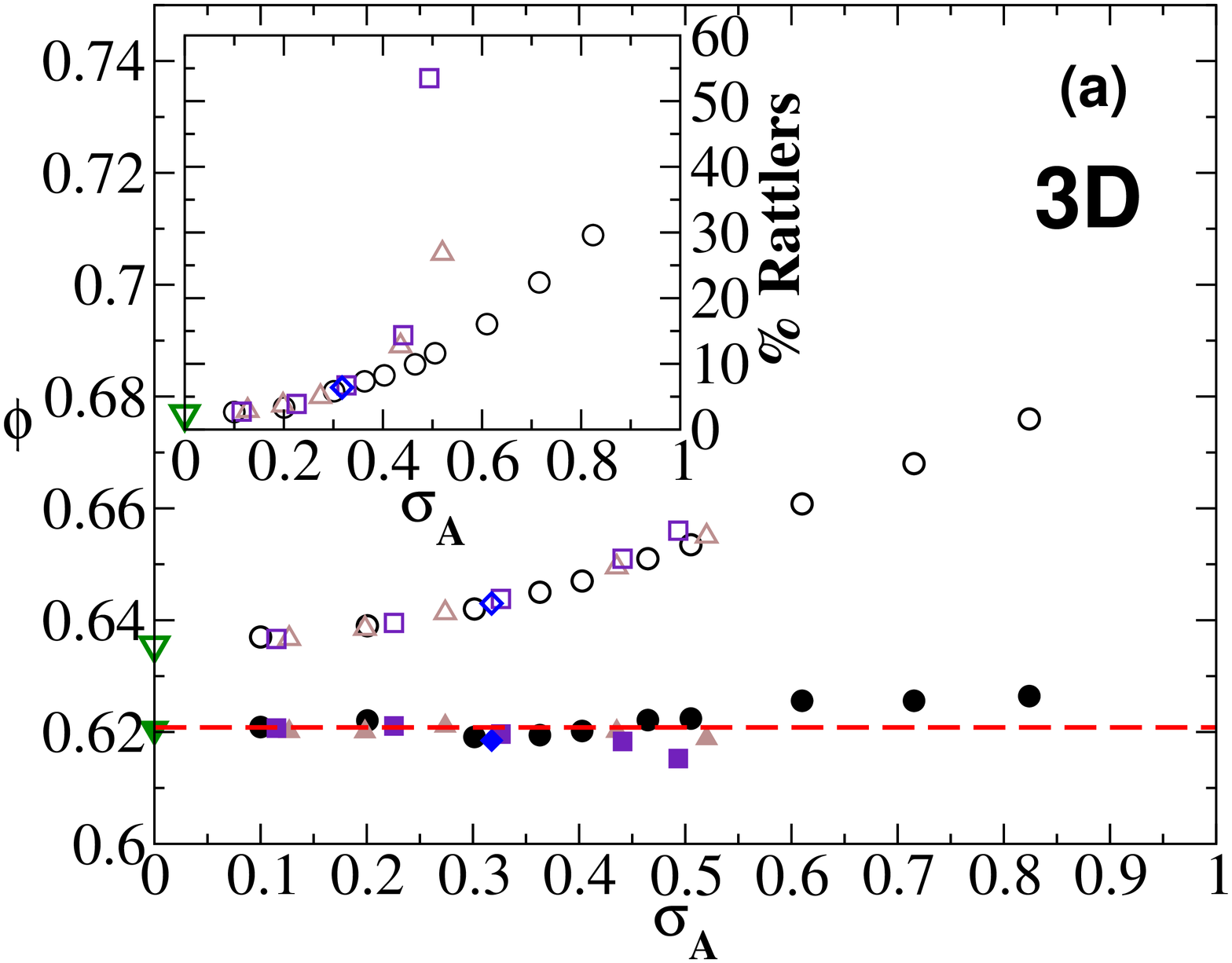}
\includegraphics[angle=270,width=68.0mm]{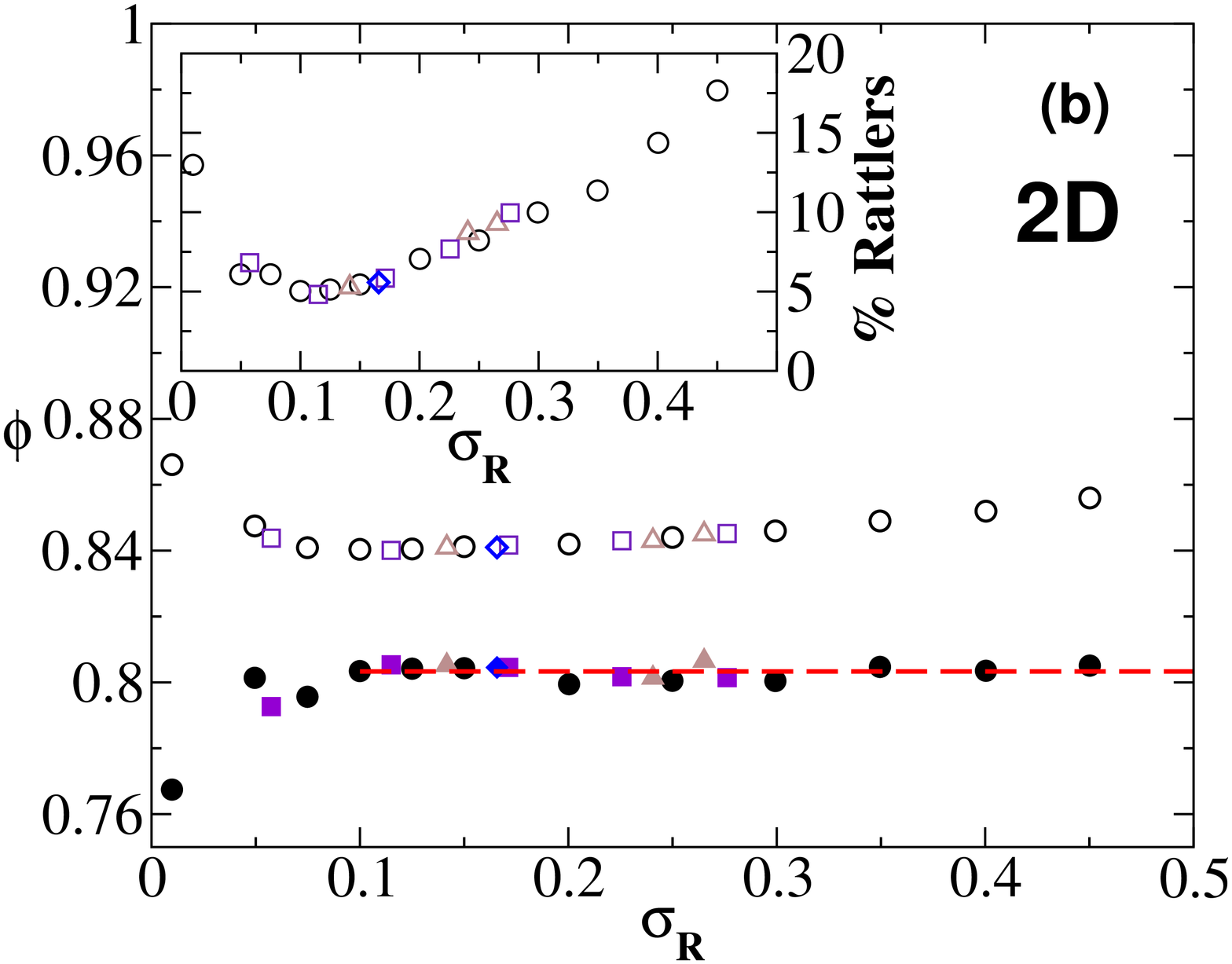}
\caption{\label{figphiRCP} (a) $\phi_{RCP}$ versus the standard deviation of the surface area distribution $\sigma_A$ for a variety of size distributions in three dimensions: (\textcolor{OliveGreen}{$\triangledown$}) monodisperse; (\textcolor{blue}{$\lozenge$}) bidisperse; (\textcolor{Purple}{$\square$}) uniform; (\textcolor{brown}{$\vartriangle$}) Gaussian; (\textcolor{black}{$\bigcirc$}) lognormal. $\phi_{RCP}$ including all particles is plotted with open symbols while solid symbols correspond to $\phi_{RCP}$ with rattlers omitted. The dashed line indicates the average $\phi_{RCP}$ with rattlers omitted for all size distributions. Inset: Percentage of rattlers at $\phi_{RCP}$ in three dimensions versus the standard deviation of the surface area distribution $\sigma_A$. (b) $\phi_{RCP}$ versus the standard deviation of the radius distribution in two dimensions with (open symbols) and without rattlers (closed symbols). The dashed line is the average $\phi_{RCP}$ with rattlers omitted for all size distributions with $\sigma_R \geq 0.1$. Inset: Percentage of rattlers at $\phi_{RCP}$ in two dimensions and the standard deviation of the radius distribution $\sigma_R$. The symbols correspond to the same data as in Figure \ref{figphiRCP}(a). }
\end{figure}

\subsection{Properties of Polydisperse Packings}

Next we define $\sigma_R$ as the normalised standard deviation of the $P(R)$ distribution where rattlers have been removed,
\begin{equation}
\sigma_R = \sqrt{ \frac{\avg{R^{2}}}{\avg{R}^2} -1}.
\end{equation}
It was also found useful for packings in three dimensions to define the standard deviation of the corresponding normalised surface area distribution as
\begin{equation}
\sigma_A = \sqrt{\frac{\avg{R^{4}}}{\avg{R^2}^2} -1}.
\end{equation}
The size distribution affects the packing density at which the isostatic point is reached\cite{Schaertl}. As the width of the size distribution is increased, $\phi_{RCP}$ becomes larger because smaller particles are able to fit between the interstices of larger particles in contact as seen in Figure \ref{figphiRCP}(a) for three dimensions and in Figure \ref{figphiRCP}(b) for two dimensions.  This also results in an increase of rattlers\cite{GrootRCP}, as shown in the insets of Figures \ref{figphiRCP}(a) and (b). $\phi_{RCP}$ only depends strongly on the width but not the shape of the size distribution. As the size distribution becomes wider the percentage of rattlers increases. For polydispersities with a large population of small particles such as the uniform distribution this results in an increase of rattlers of up to $50\%$, though $\phi_{RCP}$ is only slightly affected.

\begin{figure}
\begin{center}
\includegraphics[angle=270,width=120.0mm]{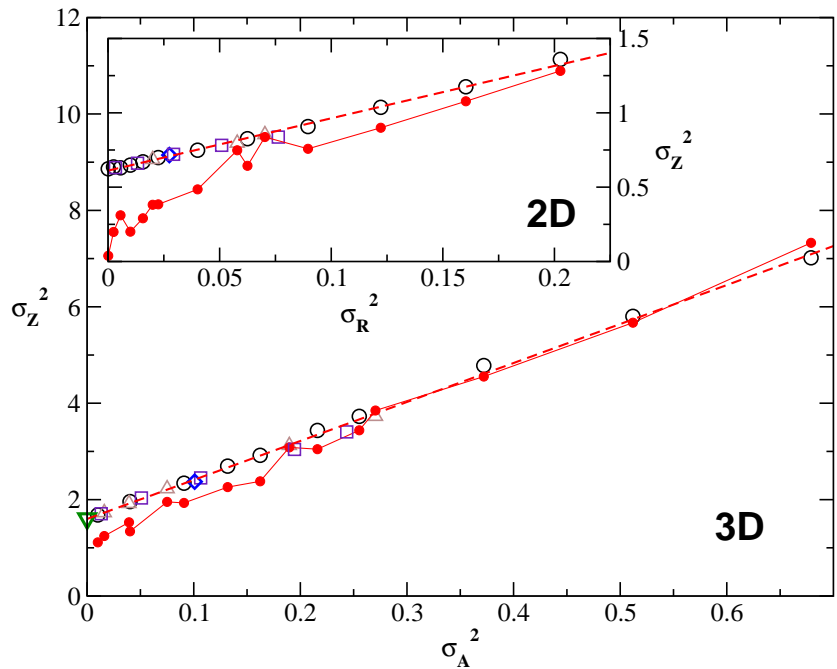}
\caption{\label{figmu2_vr} Variance of the  contact number distribution $\sigma_{Z}^{2}$ versus the variance of the area distribution $\sigma_{A}^{2}$. The dashed line corresponds to a linear fit to the data: $\sigma_{Z}^{2}$ = 1.60 + 8.09$\sigma_{A}^{2}$. The closed symbols are the predictions from Equation \ref{eqsz2sa2pred}.  Plotted in the inset is the variance of $\sigma_{Z}^{2}$ versus the radius distribution $\sigma_{R}^{2}$ for two dimensional packings. The dashed line corresponds to a linear fit to the data: $\sigma_{Z}^{2}$ = 0.61 + 3.52$\sigma_{R}^{2}$. The closed symbols are the predictions from Equation \ref{eqsz2sr2pred}. The data is labeled as in Figure \ref{figphiRCP}(a).}
\end{center}
\end{figure}

Also plotted in Figure \ref{figphiRCP}(a) and Figure \ref{figphiRCP}(b) is $\phi_{RCP}$ when the volume of rattlers is excluded. This $\phi_{RCP}$ with rattlers omitted is found to be a constant that is independent of the size distribution in three dimensions, where the average $\phi_{RCP}$ with rattlers omitted is $0.621 \pm 0.003$.  In two dimensions for $\sigma_R \geq 0.1$ the $\phi_{RCP}$ with rattlers omitted is also constant and independent of polydispersity with the average $\phi_{RCP}$ with rattlers omitted equal to $0.803 \pm 0.002$. Two dimensional disc packings with $\sigma_R < 0.1$ partially crystallise \cite{HardDiskCrystal} which leads to the increase in in $\phi_{RCP}$ and rattlers for packings in two dimensions with $\sigma_R < 0.1$.

Changing the polydispersity also affects the contact properties. As shown in Figure \ref{figmu2_vr}, changing the width of the size distribution affects the variance of the contact number distribution $\sigma_{Z}^{2}$. The standard deviation of the contact number distribution $\sigma_{Z}$ is defined as $\sigma_{Z} = \sqrt{\avg{z^{2}} - \avg{z}^{2}}.$
Broader size distributions results in broader contact number distributions. This trend is independent of the type of size distribution in both two and three dimensions. We find that $\sigma_{Z}^{2}$ increases linearly with $\sigma_{R}^{2}$ in two dimensions and $\sigma_{A}^{2}$ in three dimensions. For two dimensional cellular structures a corresponding relationship between the standard deviation of the size distribution and the standard deviation of the number of cell faces has been observed\cite{Durand, TopGeomDisorder, MikliusHilgenfeldt}. While the width of the contact number distribution is set by the width of the size distribution only, its shape does depend on the particular size distribution as can be seen in Figure \ref{figpred}.

\section{Local correlations in polydisperse packings}
\subsection{Contact number and size correlations in three dimensions}

While a large body of literature on random packings is devoted to the bulk properties of mono- and bi-disperse packings near the jamming transition \cite{Epitome,WyartRev,PackRevVanHecke,KatgertJamGeom}, important results on the local structure in polydisperse packings have emerged only in recent years \cite{GranocentricN,GranocentricSM,BrujicMeanCells}. The pioneering work by Clusel et al. established a link between the size distribution and the local structure of the packing. We have expanded on this work by investigating how the correlations between particle size and contact number depend upon the polydispersity of the packing.

\begin{figure}
\begin{center}
\includegraphics[width=100.0mm]{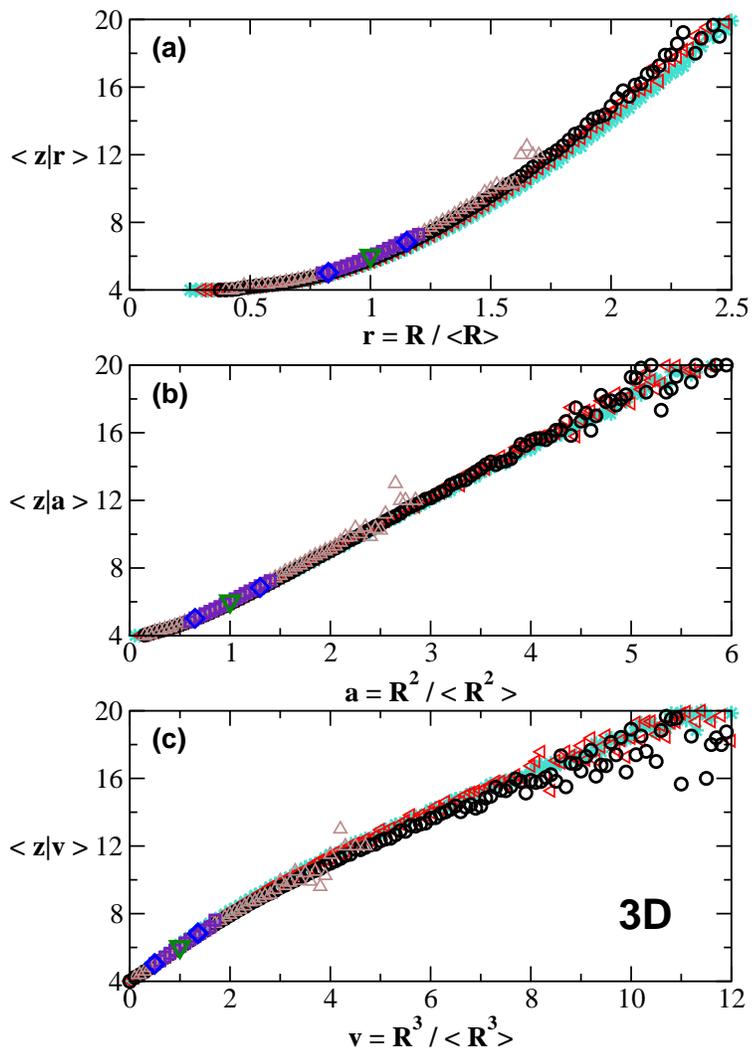}
\caption{\label{figlewis} The average of the contact number distribution for a particle of a given size for the six different size distributions at $\phi_{RCP}$: (\textcolor{OliveGreen}{$\triangledown$}) monodisperse; (\textcolor{blue}{$\lozenge$}) bidisperse, radius ratio 1:1.4; (\textcolor{Purple}{$\square$}) uniform $\sigma_A = 0.23$; (\textcolor{brown}{$\vartriangle$}) Gaussian $\sigma_A = 0.27$; (\textcolor{black}{$\bigcirc$}) lognormal $\sigma_A = 0.40$; (\textcolor{red}{$\vartriangleleft$}) lognormal $\sigma_A = 0.61$; (\textcolor{cyan}{$\ast$}) lognormal $\sigma_A = 0.72$. We present three different scalings: (a) in terms of the normalised radius $r$; (b) in terms of the normalised area $a$; (c)  in terms of the normalised volume $v$. The data are plotted over a range that illustrates the quality of the collapse.}
\end{center}
\end{figure}

\begin{figure}
\begin{center}
\includegraphics[angle=270,width=120.0mm]{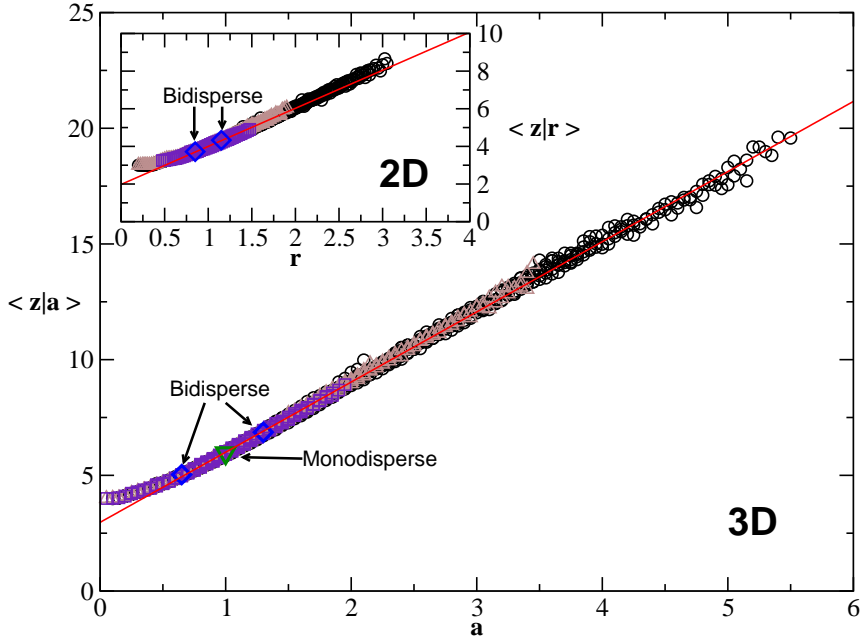}
\caption{\label{figlewis_a} The average contact number for particles of a given area $a$ at $\phi_{RCP}$ in three dimensions for all (\textcolor{OliveGreen}{$+$}) monodisperse, (\textcolor{blue}{$\lozenge$}) bidisperse, (\textcolor{Purple}{$\square$}) uniform, (\textcolor{brown}{$\vartriangle$}) Gaussian and (\textcolor{black}{$\bigcirc$}) lognormal size distributions at all the widths $\sigma_A$ we have considered (see Figure \ref{figmu2_vr}). The solid red line is a fit to Equation (\ref{eqzafit}). Inset: the average contact number for particles of a given radius $r$ in two dimensions at $\phi_{RCP}$. The solid red line is a fit to Equation (\ref{eqzrfit}). }
\end{center}
\end{figure}

The average contact number for particles of a given size is defined as
\begin{equation}
\avg{z|x} = \sum\limits_{z} z P (z | x),
\end{equation}
where $P(z|x)$ is the contact number distribution for particles of a given size $x$. The average contact number for particles of given size $x$ at $\phi_{RCP}$ is plotted for a wide range of size distributions of different widths and shape. We scaled the data in three different ways, in terms of the normalised radius, normalised surface area and normalised volume in Figure \ref{figlewis}(a), (b) and (c) respectively. In the three scalings $\avg{z|x}$ for all size distributions and polydispersities follow similar trends. Namely, larger particles have more contacts on average. This can be explained in the context of the granocentric model \cite{GranocentricN,GranocentricSM} which stipulates that larger particles have more solid angle available to accommodate neighbouring spheres. Surprisingly, these correlations are independent of polydispersity. The best collapse is observed when the scaling is in terms of the normalised area
\begin{equation}
x = a = \frac{R^{2}}{\avg{R^{2}}},
\end{equation}
as shown in Figure \ref{figlewis}(b). This collapse of the data is well described by a linear fit
\begin{equation}\label{eqzafit}
\avg{z|a} = \avg{z} + \gamma (a -1),
\end{equation}
which is plotted in Figure \ref{figlewis_a}.
The form of Equation (\ref{eqzafit}) ensures that the isostatic constraint \begin{equation}
\label{eq3Dalpha}
\avg{z} = \int\limits_{0}^{\infty}\avg{ z | a } P(a)da = 6,
\end{equation}
is satisfied. The fitting parameter is found to be $\gamma = 3.032 \pm 0.004$.
The contact number average $\avg{z|a}$ for the discrete distributions (monodisperse, bidisperse) has the same value as that of a particle of the same size in the continuous distributions. Figure \ref{figlewis}(b) shows that at $\phi_{RCP}$, the relationship between $z$ and $a$ is universal and independent of size distributions. This suggests that the local contact properties of a particle only depends upon its surface area.
This result is similar to that observed in two dimensional disc packings \cite{MobiusMe}, which will be discussed in Section \ref{2Dpred}. 

\begin{figure}
\begin{center}
\includegraphics[width=100.0mm]{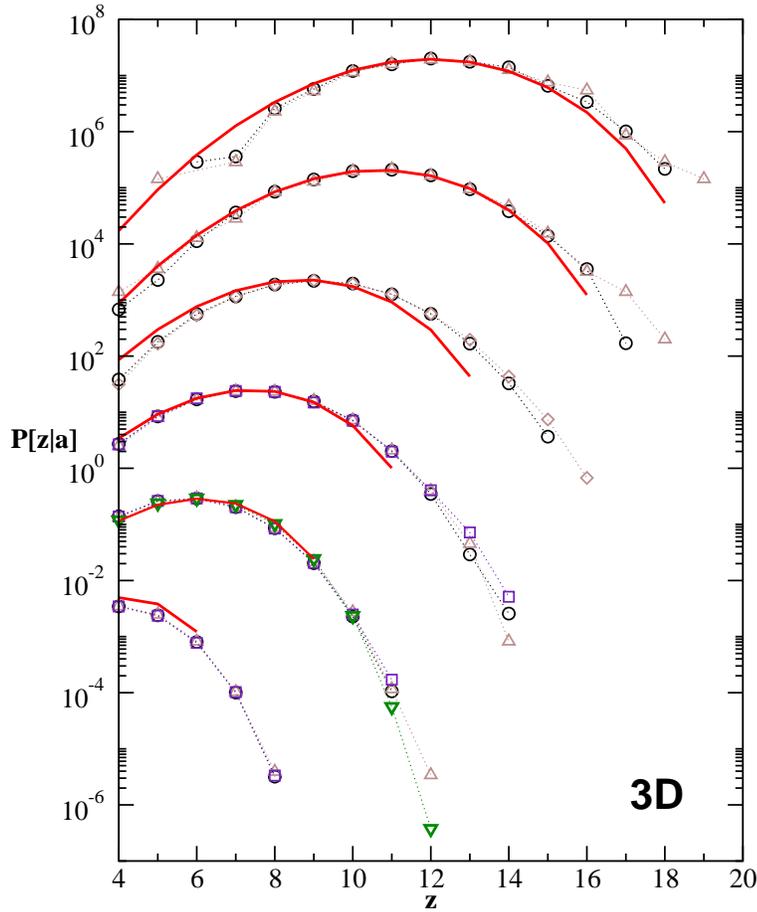}
\caption{\label{figPzapred} The contact number distribution for particles of a given size $P(z|a)$ for four size distributions at $\phi_{RCP}$ in three dimensions: (\textcolor{OliveGreen}{$\triangledown$}) monodisperse; (\textcolor{Purple}{$\square$}) uniform $\sigma_A = 0.44$; (\textcolor{brown}{$\vartriangle$}) Gaussian $\sigma_A = 0.44$; (\textcolor{black}{$\bigcirc$}) lognormal $\sigma_A = 0.40$. Plotted here is a selection of the $P(z|a)$ for 6 different intervals $a$ with the $P(z|a)$ shifted for clarity. Plotted in order of lowest to highest is $0.475< a < 0.525$; $0.975 < a < 1.025$; $1.475 < a < 1.525$; $1.975 < a < 2.025$; $2.475 < a < 2.525$; $2.975 < a < 3.025$. The solid red line is the model prediction of $P(z|a)$ from Equation (\ref{eqMOD}).}
\end{center}
\end{figure}

It must be noted that the average $\avg{z | a}\ge 4$ since we omit rattlers from our analysis. Also, for large values of $a$ the scatter in $\avg{z|a}$ is much larger due to lower statistics. For all equations fitted and figures plotted with the exception of Figures \ref{figlewis}, \ref{figPzapred}, \ref{figpred}, \ref{figlewis2D}, \ref{figPzr}, \ref{figpz2d}, data binned with less than 100 particles are omitted.

In Figure \ref{figPzapred}, a number of different contact number distributions $P(z|a)$ are plotted for given intervals of $a$. This figure demonstrates that the $P(z|a)$ distributions are independent of shape and width of the size distribution. This confirms what is suggested in Figure \ref{figlewis_a} - namely that the contact number distribution for a particle in a packing at $\phi_{RCP}$ does not depend on the global size distribution of the packing but on the size of the particle only.

\subsection{Mean field granocentric model in three dimensions}
We have shown the contact number distributions $P(z|a)$ for particles of a given size do not depend on the global size distribution of the packing but only on the size of the particle in question. This result allows us to formulate a mean field granocentric model that is similar in spirit to the one by Newhall et al. \cite{NewhallMFC} who investigated size-topology relations in tessellated packings.

Here we use a mean field approach that allows us to predict the correlations between size and contact number. In contrast to the original granocentric model \cite{GranocentricN} we explicitly exclude rattlers since their contact number is ill-defined.

Since local correlations are independent of the size distributions, we consider a particle of a given radius $R_c$ and then make a mean field assumption that all the particles surrounding it are of average radius $\avg{R}$. If this particle of size $R_c$ is in contact with another particle $\avg{R}$ it will subtend a solid angle $\Omega$ of the central particle, which is given by

\begin{equation}
\label{eqRSA}
\Omega(R_c,\avg{R}) = 2\pi\left(1 - \frac{1}{1 + \frac{\avg{R}}{R_c}}\sqrt{1 + \frac{2\avg{R}}{R_c}}\right).
\end{equation}
Since we have shown that the proper scaling of correlations between size and contact number is in terms of $a$ we rewrite Equation (\ref{eqRSA}) accordingly with all contacting particles now assumed to have an average radius $\sqrt{\avg{R^2}}$:
\begin{equation}
\label{eqASA}
\Omega(a) = 2\pi\left(1 - \frac{\sqrt{a}}{1 + \sqrt{a}}\sqrt{1 + \frac{2}{\sqrt{a}}}\right).
\end{equation}
Having obtained a value of the solid angle subtended by a single contact, the maximum number of contacts is simply
\begin{equation}
Z_{max}(a) = \frac{4\pi}{\Omega(a)}.
\end{equation}
A correction must be made to $Z_{max}$ to account for the interstices, similar to the familiar sphere kissing problem for monodisperse spheres where only $12$ spheres can be in contact with a central sphere even though there is sufficient solid angle to fit $14$ spheres\cite{Kissing}.

A prefactor $\alpha$ is introduced into the model to limit the maximum number of contacts:
\begin{equation}
\label{eqzmax}
Z_{max}(a) = \frac{2\alpha}{1 - \frac{\sqrt{a}}{1 + \sqrt{a}}\sqrt{1 + \frac{2}{\sqrt{a}}}}.
\end{equation}
In order to recover the known result of the kissing problem for monodisperse spheres, the value of $\alpha$ would have to be $0.8708$. In our model, however, the value of $\alpha$ will turn out to be less than that due to additional constraints.

Following the granocentric approach \cite{GranocentricN}, we now make an ansatz that the distribution of the number of particles in contact with a particle of size $a$ is given by a binomial distribution.
\begin{align}\label{eqBIO}
P(z|a)&=B(z;Z_{max}(a),p),\\
&=\frac{Z_{max}!}{z!(Z_{max}-z)!}p^z(1-p)^{Z_{max}-z},\nonumber
\end{align}
where $B(z;Z_{max}(a),p)$ is a binomial distribution with the maximum number of trials, that is the number of times in which a particle can attempted to be placed in contact with the particle of size $a$, given by $Z_{max}$ and $p$ is the acceptance probability that a particle will be in contact. The probability $p$ is the other free parameter in the model.

In order to omit rattlers we truncate the binomial distribution for $z < 4$ by including a Heaviside function $H(z-4)$ and a normalisation constant $C$, so $P(z|a)$ becomes,
\begin{align}\label{eqMOD}
P(z|a) &= B'(z;Z_{max}(a),p), \\
          &= C\frac{Z_{max}!}{z!(Z_{max}-z)!}p^z(1-p)^{Z_{max}-z}H(z-4). \nonumber
\end{align}
Note that this is in contrast to the original granocentric model which did not exclude rattlers. This allows us to make a prediction for the contact number average for a given particle size,
\begin{equation}
\label{eqzapred}
\avg{z|a}= \sum\limits_{z=4}^{Z_{max}(a)}zP(z|a),
\end{equation}
and the corresponding variance of the contact number for a given particle size
\begin{equation}
\label{eqmupred}
\avg{\sigma_{Z}^{2}|a} = \sum\limits_{z=4}^{Z_{max}(a)}(z-\avg{z|a})^2P(z|a),
\end{equation}
as well as a prediction of the global contact number distribution of the packing
\begin{equation}
\label{eqpzpred}
P(z) = \int\limits_{0}^{\infty}P(z|a)P(a)da.
\end{equation}

\begin{figure}
\begin{center}
\includegraphics[width=100.0mm]{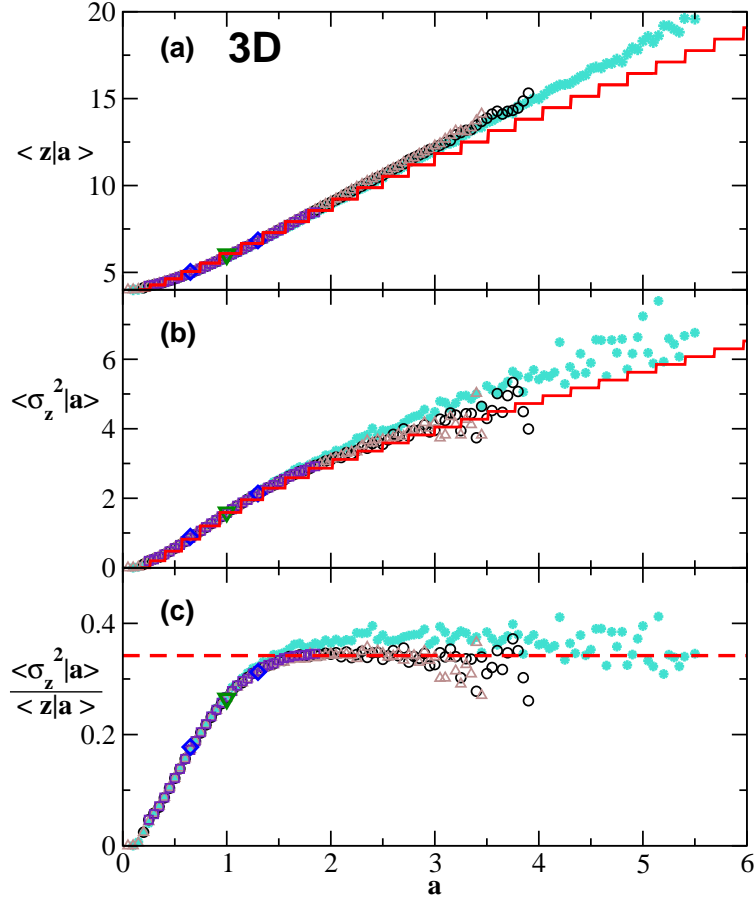}
\caption{\label{figlewis_pred} (a) The average of the contact number distribution for particle of a given size for five different size distributions at $\phi_{RCP}$. The model prediction from Equation (\ref{eqzapred}) is plotted as the solid red line. (b) The variance of the contact number distribution for particle of a given $a$. The model prediction from Equation (\ref{eqmupred})  is plotted as the solid red line. (c) The ratio of the variance to the average of the contact number distribution for particles of a given size $a$. The dashed line denotes the value of the acceptance probability $p$ as found from the data. The size distributions and symbols plotted in all three panels are the same as in Figure \ref{figPzapred} with the addition of: (\textcolor{blue}{$\lozenge$}) bidisperse, radius ratio 1:1.4 and (\textcolor{cyan}{$\ast$}) lognormal $\sigma_A = 0.72$.}
\end{center}
\end{figure}

Given a size distribution $P(a)$ and using Equation (\ref{eqMOD}), a prediction for the contact number distribution can be made for any packing at $\phi_{RCP}$. The acceptance probability $p$ can be determined through a property of the binomial distribution. If $X \sim B(n,p)$ is a random variable from a binomial distribution $B$ with $n$ trials  then the mean is given by
\begin{equation*}
E[X] = np,
\end{equation*}
and the variance is given by
\begin{equation*}
Var[X] = np(1-p).
\end{equation*}
Therefore the ratio of the variance to the mean of a binomial distribution is a constant given in terms of $p$, which in the context of our model is given by
\begin{equation}
\frac{\avg{\sigma_{Z}^{2}|a}}{\avg{z|a}} = 1-p.
\end{equation}

Equation (\ref{eqMOD}) is a truncated binomial distribution and therefore the ratio $\frac{\avg{\sigma_{Z}^{2} | a}}{ \avg{z| a}}$ plotted in Figure \ref{figlewis_pred}(c) is only expected to reach a constant at sufficiently large values of $a$ where the truncation becomes negligible. Indeed, for $a\gtrsim 2$ the ratio plateaus at $0.342\pm 0.006$ which corresponds to $p =0.658 \pm 0.006$.

After obtaining the probability $p$ directly from the data we can fix the second parameter $\alpha$ by imposing the constraint as stated in Equation (\ref{eq3Dalpha}), namely that the global average contact number of the packing $\avg{z}$ must be equal to $6$. This results in $\alpha$ taking a value of $0.625$. Surprisingly, $\alpha$ \emph{does not} depend on polydispersity, therefore the two free parameters of the model, $\alpha$ and $p$, can be fixed for all size distributions at $\phi_{RCP}$. This may be related to the fact $\phi_{RCP}$ without the rattlers, which are explicitly omitted in this model, is a constant (Figure \ref{figphiRCP}).

The constancy of the two parameters is a significant simplification to the original granocentric model, where the acceptance probability and maximum solid angle need to be determined for each polydispersity separately.

\begin{figure}
\begin{center}
\includegraphics[width=100.0mm]{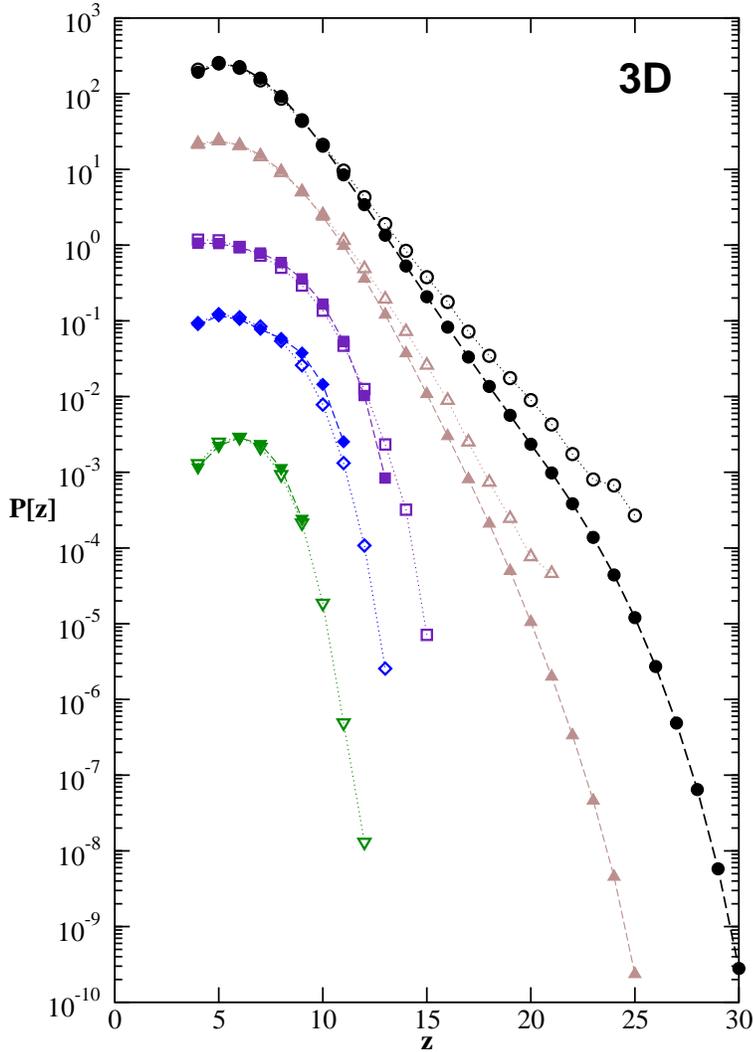}
\caption{\label{figpred} The contact number distribution for five different size distributions at $\phi_{RCP}$. The $P(z)$ are shifted for clarity. The open symbols represent the simulation data and the closed  symbols represent the model prediction from Equation (\ref{eqpzpred}). The size distributions and symbols plotted are the same as in Figure \ref{figlewis_pred}.}
\end{center}
\end{figure}

Comparing the prediction of the average $\avg{z | a }$ from Equation (\ref{eqzapred}) with the data as shown in Figure \ref{figlewis_pred}(a), we see that the model is in good agreement with the data over a large range of $a$. Only for large values of $a$ it deviates slightly. Similarly, the model prediction of the variance $\avg{\sigma_{Z}^{2} | a}$ (Equation (\ref{eqmupred})) agrees well with the data as shown in Figure \ref{figlewis_pred}(b). Note that the staircase structure exhibited by the model in Figure \ref{figlewis_pred}(a)-(b) is due to the discrete nature of the binomial distribution. Finally, we can compare the prediction for the contact number distribution $P(z)$ from Equation (\ref{eqpzpred}) with the data for a wide range of size distributions as shown in Figure \ref{figpred}. As mentioned earlier, the same parameters $\alpha$ and $p$ are used for all polydispersities.

For the continuous size distributions the agreement is excellent, with slight deviations in the tails, while for the discrete size distributions the model fails to reproduce the tails of the distribution. This is a consequence of the $\alpha$ parameter which takes on a value that limits the number of contacts to less
than the maximum number of contacts allowed by geometry.

For example, to recover the maximum number of contacts in monodisperse packings, $\alpha$ would need to be  $0.8708$. This would violate the constraint imposed by Equation (\ref{eq3Dalpha}) as the probability $p$ is set by the data and cannot be adjusted. As the size distribution becomes wider this discrepancy becomes less pronounced.

\subsection{\label{2Dpred}Mean field granocentric model in two dimensions}

\begin{figure}
\begin{center}
\includegraphics[width=100.0mm]{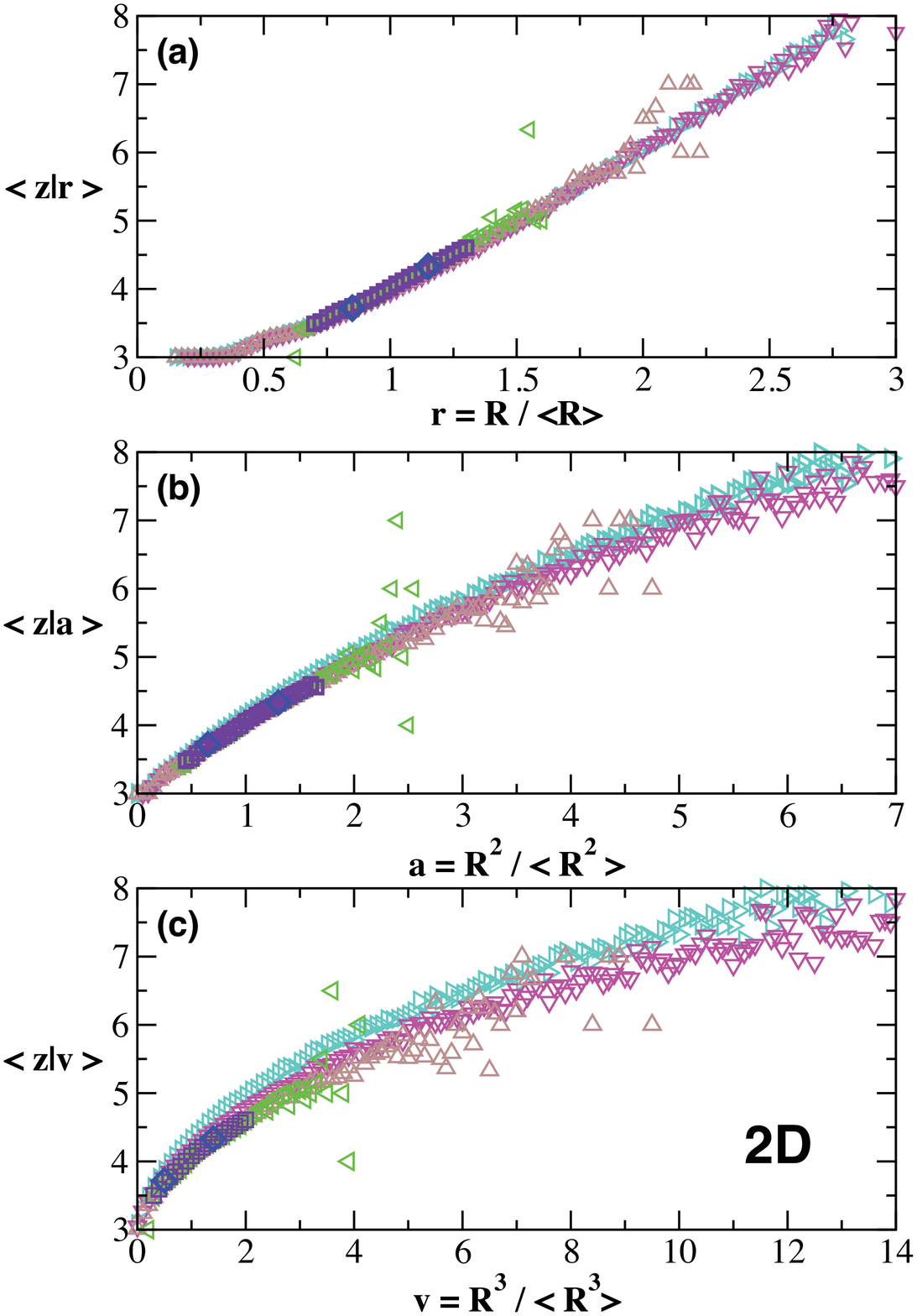}
\caption{\label{figlewis2D} The average of the contact number distribution for a particle of a given size for the six different size distributions in two dimensions at $\phi_{RCP}$: (\textcolor{blue}{$\lozenge$}) bidisperse, radius ratio 1:1.4; (\textcolor{Purple}{$\square$}) uniform $\sigma_R = 0.17$; (\textcolor{green}{$\vartriangleleft$}) lognormal $\sigma_R = 0.10$; (\textcolor{brown}{$\vartriangle$}) Gaussian $\sigma_R = 0.24$; (\textcolor{magenta}{$\triangledown$}) lognormal $\sigma_R = 0.35$; (\textcolor{cyan}{$\triangleright$}) lognormal $\sigma_R = 0.45$. We present three different scalings: (a) in terms of the normalised radius $r$; (b) in terms of the normalised area $a$; (c)  in terms of the normalised volume $v$. The data are plotted over a range that illustrates the quality of the collapse.}
\end{center}
\end{figure}

\begin{figure}
\begin{center}
\includegraphics[width=100.0mm]{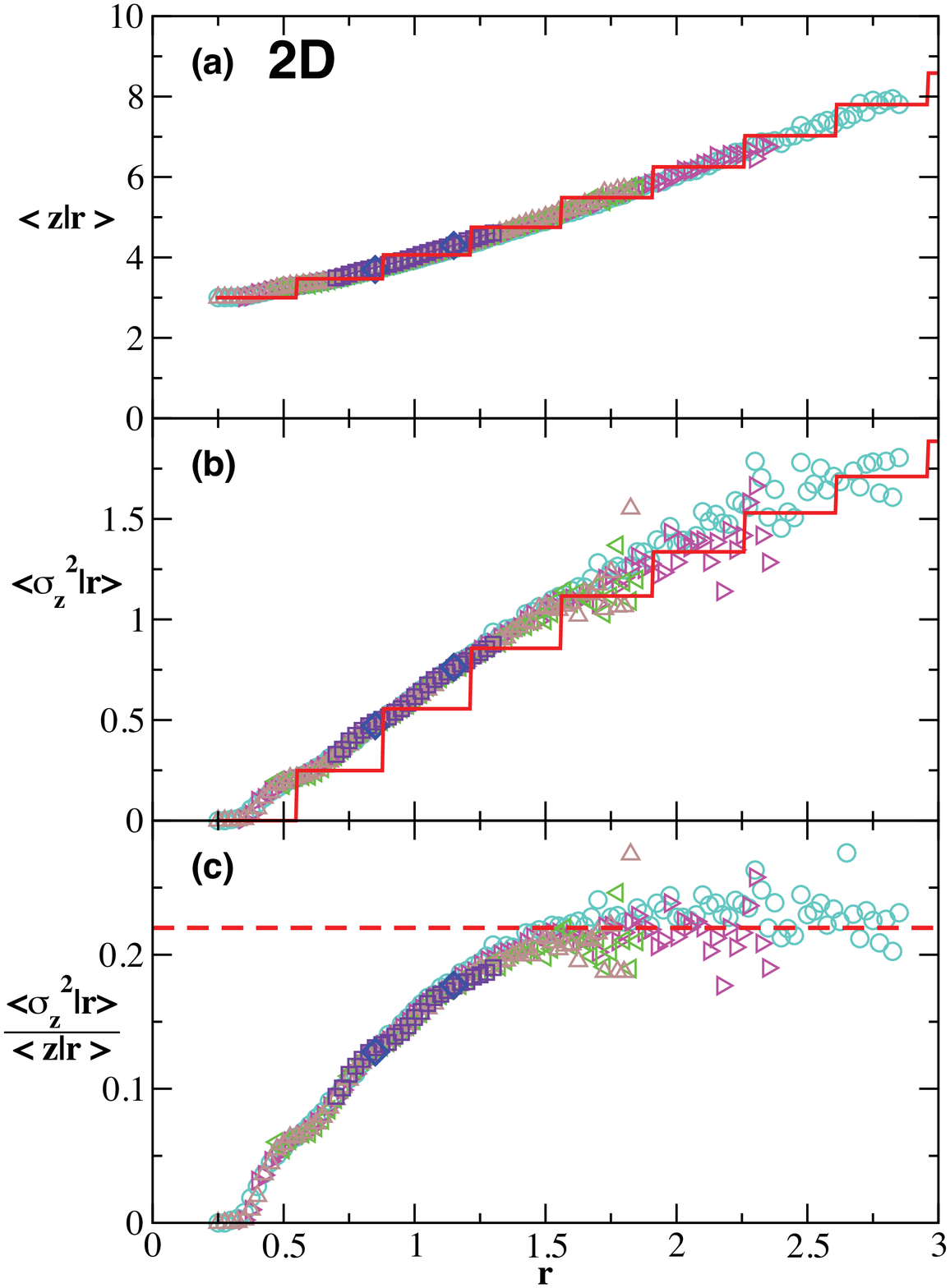}
\caption{\label{figzr}(a) The average of the contact number distribution for particles of a given radius $r$ in two dimensions for six different size distributions at $\phi_{RCP}$:(\textcolor{blue}{$\lozenge$}) bidisperse, radius ratio 1:1.4; (\textcolor{Purple}{$\square$}) uniform $\sigma_R = 0.17$; (\textcolor{brown}{$\vartriangle$}) Gaussian $\sigma_R = 0.24$; (\textcolor{green}{$\vartriangleleft$}) lognormal $\sigma_R = 0.20$; (\textcolor{magenta}{$\triangleright$}) lognormal $\sigma_R = 0.30$; (\textcolor{cyan}{$\bigcirc$}) lognormal $\sigma_R = 0.40$. The model prediction is plotted as the solid red line. (b) The variance $\avg{ \sigma_{Z}^{2}|r }$ for the same size distributions. The model prediction is plotted as the solid red line. (c) The ratio of the variance to the average of P$(z|r)$. The dashed red line denotes the value of the acceptance probability $p$ as found from the data.}
\end{center}
\end{figure}

We now use the same approach for 2D polydisperse packings. In Figure \ref{figlewis2D}(a), (b) and (c) the average contact number for particles of a given size $x$ at $\phi_{RCP}$ is plotted and scaled in terms of the normalised radius, normalised surface area and normalised volume. In the three scalings the $\avg{z|x}$ for all size distributions and polydispersities follow similar trends, namely that larger particles have more contacts on average. Similar to the three dimensional case, the best collapse of the data is found when the scaling is
\begin{equation}
x = \frac{R^{D-1}}{\avg{R^{D-1}}},
\end{equation}
as shown in Figure \ref{figlewis2D}(a). Therefore, the proper scaling variable for size-contact number correlations in 2D packings is $x=r$.
The inset of Figure \ref{figlewis_a} exhibits a similar collapse of the average contact number for particles of a given radius $\avg{z|r}$ for a wide range of size distributions to that found for $\avg{z|a}$ in three dimensions. Similar to Equation (\ref{eqzafit}), the average contact number $\avg{z|r}$ is well fit by a linear function of the form
\begin{equation}
\label{eqzrfit}
\avg{z|r} = \avg{z} + \gamma_{2D} (r -1),
\end{equation}
with the fit parameter $\gamma_{2D} = 2.023 \pm 0.007$.

In Figure \ref{figzr} we plot the two dimensional equivalent of Figure \ref{figlewis_pred}, except with the size of the particle now represented by the normalised radius $r$ instead of the normalised surface area $a$. The model prediction that appears in Figure \ref{figzr}(a) for $\avg{z|r}$ is analogous to the model outlined in the previous section. The principle adjustment is that the maximum number of discs that can be placed in contact with a disc of radius $r$ must now be expressed in terms of the available angle rather than the solid angle. Equation (\ref{eqzmax}) now reads as
\begin{equation}
Z_{max}(r) = \frac{\alpha \pi}{\sin^{-1}\left(\frac{1}{1+r}\right)},
\end{equation}
which affects the number of trials of the binomial distribution. Also rattlers, particles with less than $3$ contacts in 2D packings, are excluded from the binomial distribution.

\begin{figure}
\begin{center}
\includegraphics[width=100.0mm]{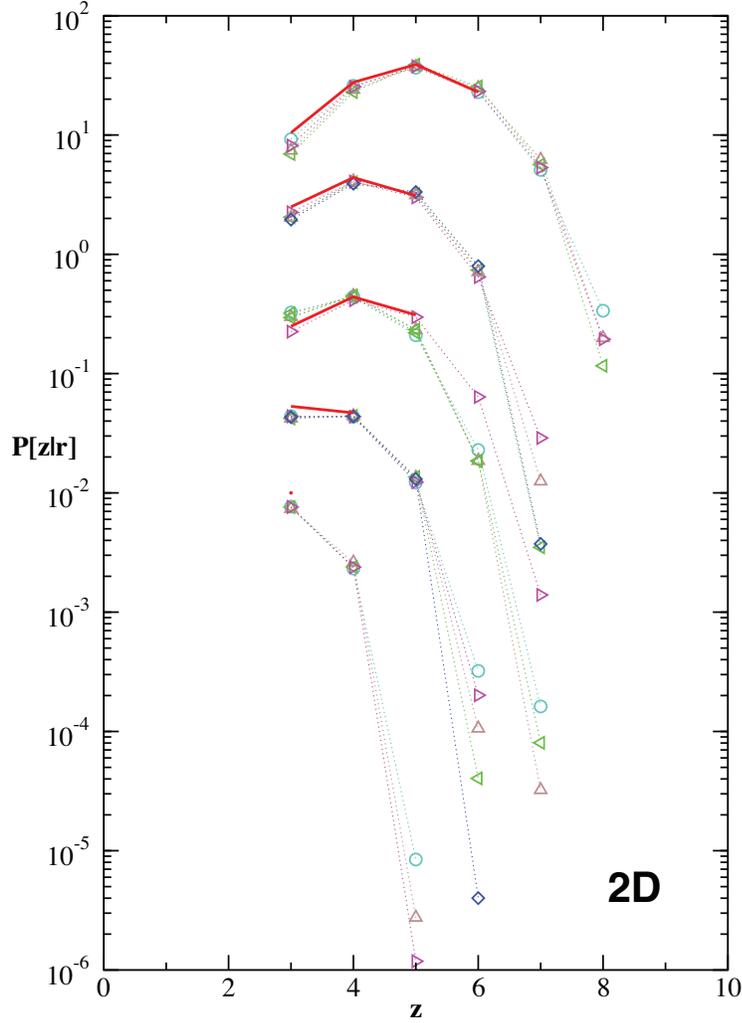}
\caption{\label{figPzr} The contact number distribution for discs of a given radius $r$, $P(z|r)$ for the same size distributions as plotted in the the inset of Figure \ref{figzr} at $\phi_{RCP}$. Plotted here is a selection of $P(z|r)$ for five different intervals $r$ with the $P(z|r)$ shifted for clarity. Plotted in order of lowest to highest is $0.475< a < 0.525$; $0.825 < a < 0.875$; $0.975 < a < 1.025$; $1.125 < a < 1.175$; $1.475 < a < 1.525$.  The solid red line is the model prediction of $P(z|r)$.}
\end{center}
\end{figure}

The acceptance probability $p$ can be determined in the same fashion using the ratio between the variance of the contact number for particles of a given radius $\avg{\sigma_{Z}^{2} | r}$ and the corresponding average $\avg{z | r}$ as shown in Figure \ref{figzr}(c). As before, the ratio appears to plateau for large particles which is consistent with a truncated binomial distribution. The value for $p$ found from the data shown in Figure \ref{figzr}(c) is $0.78 \pm 0.02$.

\begin{figure}
\begin{center}
\includegraphics[width=100.0mm]{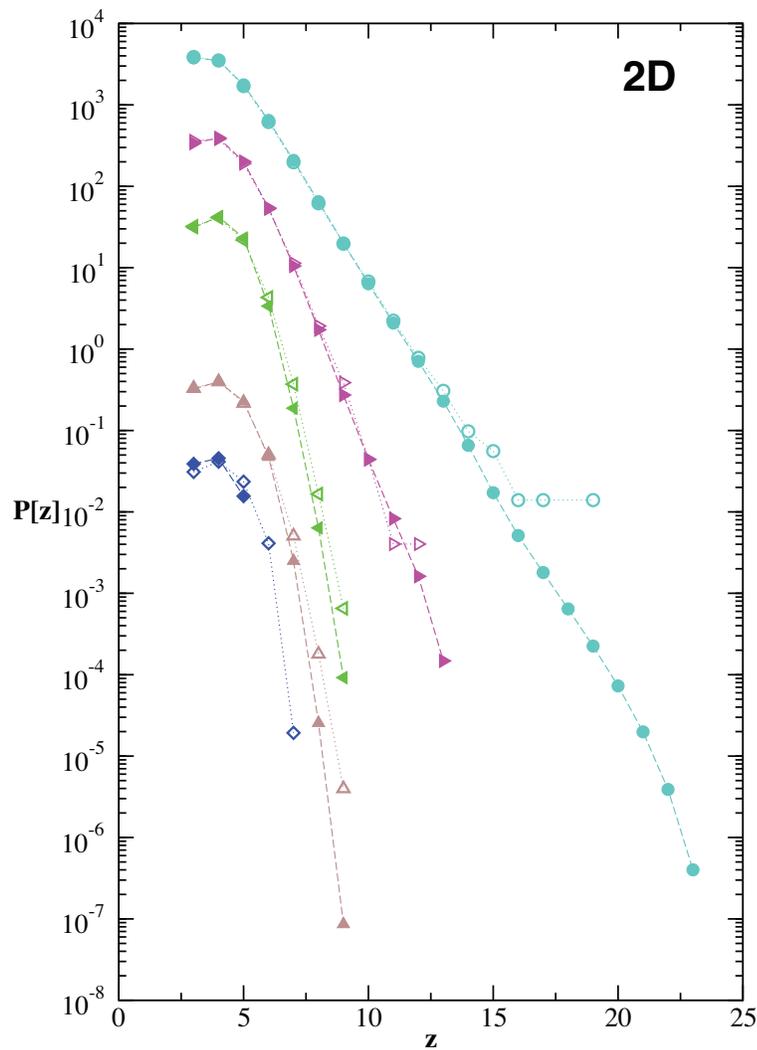}
\caption{\label{figpz2d} The contact number distribution for a number of different types of polydispersities at $\phi_{RCP}$ in two dimension. The data from simulation is plotted as open symbols and the prediction from the model is plotted as closed symbols. The parameters $p = 0.78$ and $\alpha = 0.894$ are used for all size distributions. The same size distributions are plotted with the same symbols as in Figure \ref{figzr}. Distributions are shifted for clarity.}
\end{center}
\end{figure}

Analogously to the 3D case, $\alpha$ is determined by the isostatic constraint which is
\begin{equation}
\label{eq2Dalpha}
\avg{z} = \int\limits_{0}^{\infty}\avg{z | r}P(r)dr = 4.
\end{equation}
in 2D packings with the size distribution now given in terms of $r$ rather than $a$. The value of $\alpha$ for two dimensions is found to be $0.894$. The equivalent value $\alpha$ that would recover the correct answer for the kissing problem in two dimensions is $1$.

The justification for this model is that the correlation between size and contact number is independent of of polydispersity, analogous to the results in three dimensions. Similar to Figure \ref{figPzapred} a number of difference size distributions are plotted for given intervals of $r$ in Figure \ref{figPzr}. For each $r$ interval plotted all the $P(z|r)$ collapse independent of size distribution therefore validating the basis of the model. However, Figure \ref{figPzr} highlights some of the weaknesses of the model. For example, for the lowest interval of $r$ where the data shows a range of contact numbers $z$ but the model predicts that only 3 discs can fit around a disc of that size. This discrepancy at low $r$ is due the contact limiting parameter $\alpha$. For larger values of $r$ the model prediction of $P(z|r)$ is in better agreement with the data.

The global contact distribution can then be predicted from the two dimensional equivalent of Equation (\ref{eqpzpred}),
\begin{equation}
\label{eqpzpred2d}
P(z) = \int\limits_{0}^{\infty} P(z|r)P(r)dr.
\end{equation}
Figure \ref{figpz2d} shows good agreement between the predictions and data for a wide range of size distributions. Similar to the results in three dimensions, the prediction is in closer agreement with the data for wider size distributions.

\subsection{Size of a particle with contact number $z$}
\subsubsection{Size of a particle with contact number z in three dimensions}

\begin{figure}
\begin{center}
\includegraphics[angle=270,width=120.0mm]{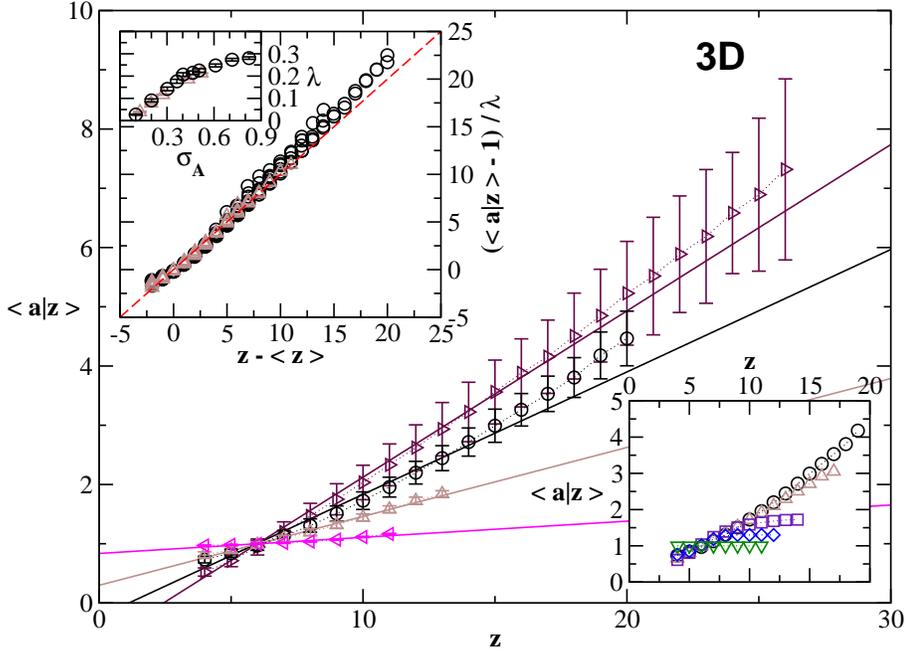}
\caption{\label{figaz}The average area of particles with a given contact number for four different size distributions in three dimensions at $\phi_{RCP}$:  (\textcolor{magenta}{$\vartriangleleft$}) lognormal $\sigma_A = 0.10$;  (\textcolor{brown}{$\vartriangle$}) Gaussian $\sigma_A = 0.27$;  (\textcolor{black}{$\bigcirc$}) lognormal $\sigma_A = 0.40$;  (\textcolor{Maroon}{$\vartriangleright$}) lognormal $\sigma_A = 0.82$. The solid lines are fits to Equation \ref{eqaz}. Top inset shows the average area of particles with a given contact number for all (\textcolor{black}{$\bigcirc$}) lognormal and (\textcolor{brown}{$\vartriangle$}) Gaussian size distributions at $\phi_{RCP}$, collapsed after rescaling by fitting the data to Equation (\ref{eqaz}). The dashed red line corresponds to a slope of 1. Inset of the top inset shows the fit parameter $\lambda$ as a function of $\sigma_A$. Bottom inset shows relationship between $\avg{a|z}$ versus $z$ for the same size distributions as plotted in Figure \ref{figlewis_pred} using the same symbols.}
\end{center}
\end{figure}

We have shown that $\avg{z|a}$, the average contact number for a particle of a given size is independent of the size distribution. However, it is important to emphasise that the converse is not true. In general, $\avg{a|z}$, the average area of particles that have $z$ contacts,  which is defined as,
\begin{equation}
\avg{a|z} = \int_0^{\infty} a P(a|z) da,
\end{equation}
is not equal to $\avg{z|a}$. The bottom inset of Figure \ref{figaz} shows that $\avg{a|z}$ is not independent of size distribution. 

Continuous size distributions, lognormal and Gaussian are well approximated by a linear relationship
\begin{equation}
\label{eqaz}
\avg{a|z} = 1 + \lambda (z - \avg{z}).
\end{equation}
This functional form ensures that $\sum_z \avg{a|z}P(z)=1$. When rescaled by the fitting parameter $\lambda$, $\avg{a|z}$  collapses for all lognormal and Gaussian size distributions, as shown in the top inset of Figure \ref{figaz}. While the overall trend of $\avg{a|z}$ for lognormal and Gaussian distributions is linear there are deviations. 

Size distributions that lack tails, such as the discrete distributions, have a different functional form of $\avg{a|z}$ because of the large population of big spheres that can take a wide range of $z$ as seen in Figure \ref{figPzapred}. This causes $\avg{a|z}$ to plateau at large $z$, as shown in the bottom inset of Figure \ref{figaz}. In the limit of monodisperse packings, $\avg{a|z}=1$, which corresponds to $\lambda=0$.

The linear relationship between size and contact number is similar to Lewis' law \cite{Lewis} for two dimensional cellular structures. 

The two different relationships between $a$ and $z$ arise from being calculated from two different conditional probabilities, the discrete distribution $P(z|a)$ and the continuous distribution $P(a|z)$, which are related by Bayes Theorem,
  \begin{equation}
  \label{eqBayes}
    P(z|a)=P(a|z)\frac{P(z)}{P(a)}.
  \end{equation}
Therefore, $P(z|a)$ and $P(a|z)$, in addition to being discrete and continuous distribution respectively, are related but not the same. Hence the fitting parameter $\lambda$ is not simply the inverse of $\gamma$.

From Equation (\ref{eqBayes}), the two conditional averages $\avg{z|a}$ and $\avg{a|z}$ can be related by
\begin{equation}
\label{eqazeq}
 \int_0^{\infty} a \avg{z|a} P(a) da = \sum\limits_{z=4}^{\infty}z \avg{a|z}P(z).
\end{equation}
Then substituting the linear fits of Equation (\ref{eqzafit}) and Equation (\ref{eqaz}) into Equation (\ref{eqazeq}) a relationship can be found between the width of the size distribution and the width of the contact number distribution,
\begin{equation}
\label{eqsz2sa2pred}
\sigma_{Z}^{2} = \frac{\gamma}{\lambda(\sigma_{A}^2)}\sigma_{A}^2.
\end{equation}
While $\gamma$ is a constant, $\lambda$ clearly depends on the width of the size distribution $\sigma_A$ as shown in the inset of Figure \ref{figaz}. Although the functional form of $\lambda(\sigma_A)$ is unclear, we can substitute the values for $\lambda$ into Equation (\ref{eqsz2sa2pred}) to compare with the data for $\sigma_{Z}^{2}$ versus $\sigma_{A}^{2}$ from Figure \ref{figmu2_vr}. The agreement is good for broad distributions but less accurate for narrow ones where $\avg{a|z}$ is not well approximated by the linear fit (Equation \ref{eqaz}).

\subsubsection{Size of a particle with contact number z in two dimensions}

\begin{figure}
\begin{center}
\includegraphics[angle=270,width=120.0mm]{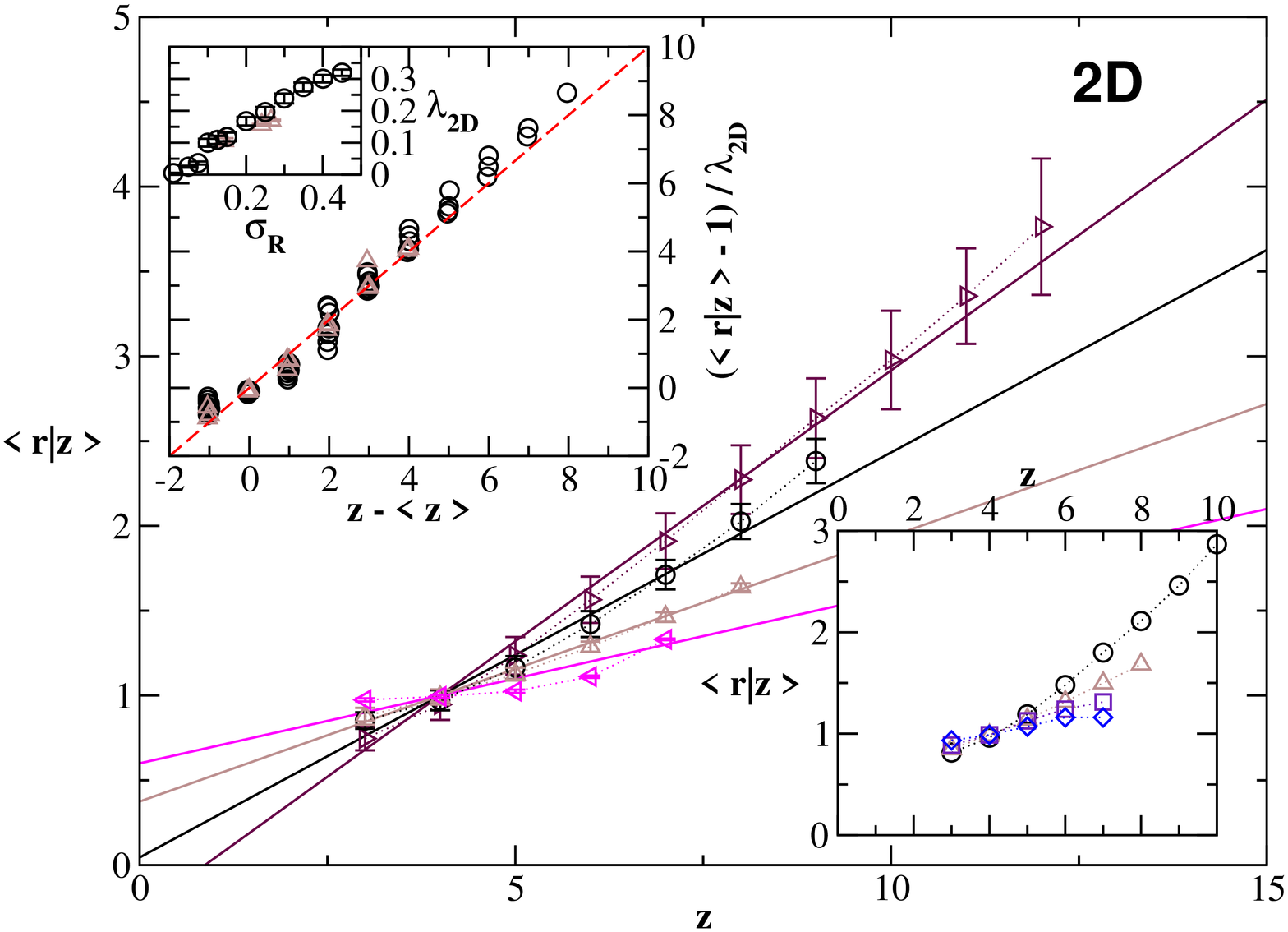}
\caption{\label{figrz}The average radius of particles with a given contact number for four different size distributions in two dimensions at $\phi_{RCP}$:  (\textcolor{magenta}{$\vartriangleleft$}) lognormal $\sigma_R = 0.10$;  (\textcolor{brown}{$\vartriangle$}) Gaussian $\sigma_R = 0.24$;  (\textcolor{black}{$\bigcirc$}) lognormal $\sigma_R = 0.30$;  (\textcolor{Maroon}{$\vartriangleright$}) lognormal $\sigma_R = 0.45$. The solid lines are fits to Equation \ref{eqrz}. Top inset shows the average area of particles with a given contact number for all (\textcolor{black}{$\bigcirc$}) lognormal and (\textcolor{brown}{$\vartriangle$}) Gaussian size distributions at $\phi_{RCP}$, collapsed  after rescaling by fitting the data to Equation (\ref{eqrz}). The dashed red line corresponds to a slope of 1. Inset of the top inset shows the fit parameter $\lambda$ as a function of $\sigma_A$. Bottom inset shows relationship between $\avg{a|z}$ versus $z$ for the size distributions: (\textcolor{blue}{$\lozenge$}) bidisperse; (\textcolor{Purple}{$\square$}) uniform $\sigma_R = 0.23$;  (\textcolor{brown}{$\vartriangle$}) Gaussian $\sigma_R = 0.27$;  (\textcolor{black}{$\bigcirc$}) lognormal $\sigma_R = 0.35$.}
\end{center}
\end{figure}

Similar correlations are also observed in two dimensions. By defining the average radius for particles with a given contact number $\avg{r|z}$ as,
\begin{equation}
\avg{r|z} = \int_0^{\infty} r P(r|z) dr,
\end{equation}
we find that $\avg{r|z}$ is well approximated by a linear relationship similar to Equation (\ref{eqaz}):
\begin{equation}
\label{eqrz}
\avg{r|z} = 1 + \lambda_{2D} (z - \avg{z}),
\end{equation}
The fits to this equation shown in Figure \ref{figrz} agree well for the 2D Gaussian and lognormal disc distributions. Analogously to the 3D packings, $\avg{r|z}$ for size distributions without tails shows deviations from the linear fit at large $z$ as shown in the bottom inset of Figure \ref{figrz}, though it is less pronounced due to the smaller range of contact numbers in 2D packings. 

In the top inset of Figure \ref{figrz} the correlations $\avg{r|z}$ for all lognormal and Gaussian size distributions are rescaled to highlight the agreement with Equation (\ref{eqrz}). The fitting parameter, $\lambda_{2D}$ is shown in the inset of Figure \ref{figrz} as a function of the polydispersity $\sigma_R$.

Analogously to the 3D case, a relationship between the standard deviations of the size distribution and the contact number distribution can be written assuming that size-contact number correlations are approximately linear,
\begin{equation}
\label{eqsz2sr2pred}
\sigma_{Z}^{2} = \frac{\gamma_{2D}}{\lambda_{2D}(\sigma_{R}^2)}\sigma_{R}^2.
\end{equation}
Comparison of this relation with the data is shown in the inset of Figure \ref{figmu2_vr}. While this relation works well for broad distributions, narrow ones are not captured well for the same reasons as in the 3D case. Also, in 2D packings partial crystallization takes place for narrow size distributions.
\section{Nearest neighbour correlations in packings}

\subsection{Nearest neighbour contact number correlations}

In the previous section we proposed a mean-field model based on the assumption that the packing is spatially uncorrelated. Specifically, we assume that the contact number and size of a particle is uncorrelated to the contact number and size of its contacting neigh ours. This assumption is implicitly made in many recent models that predict the density\cite{SongMakse}, contact distribution\cite{GranocentricN, NewhallMFC} and force networks\cite{TigheForceNet} in disordered packings. 

\begin{figure}
\begin{center}
\includegraphics[width=130.0mm]{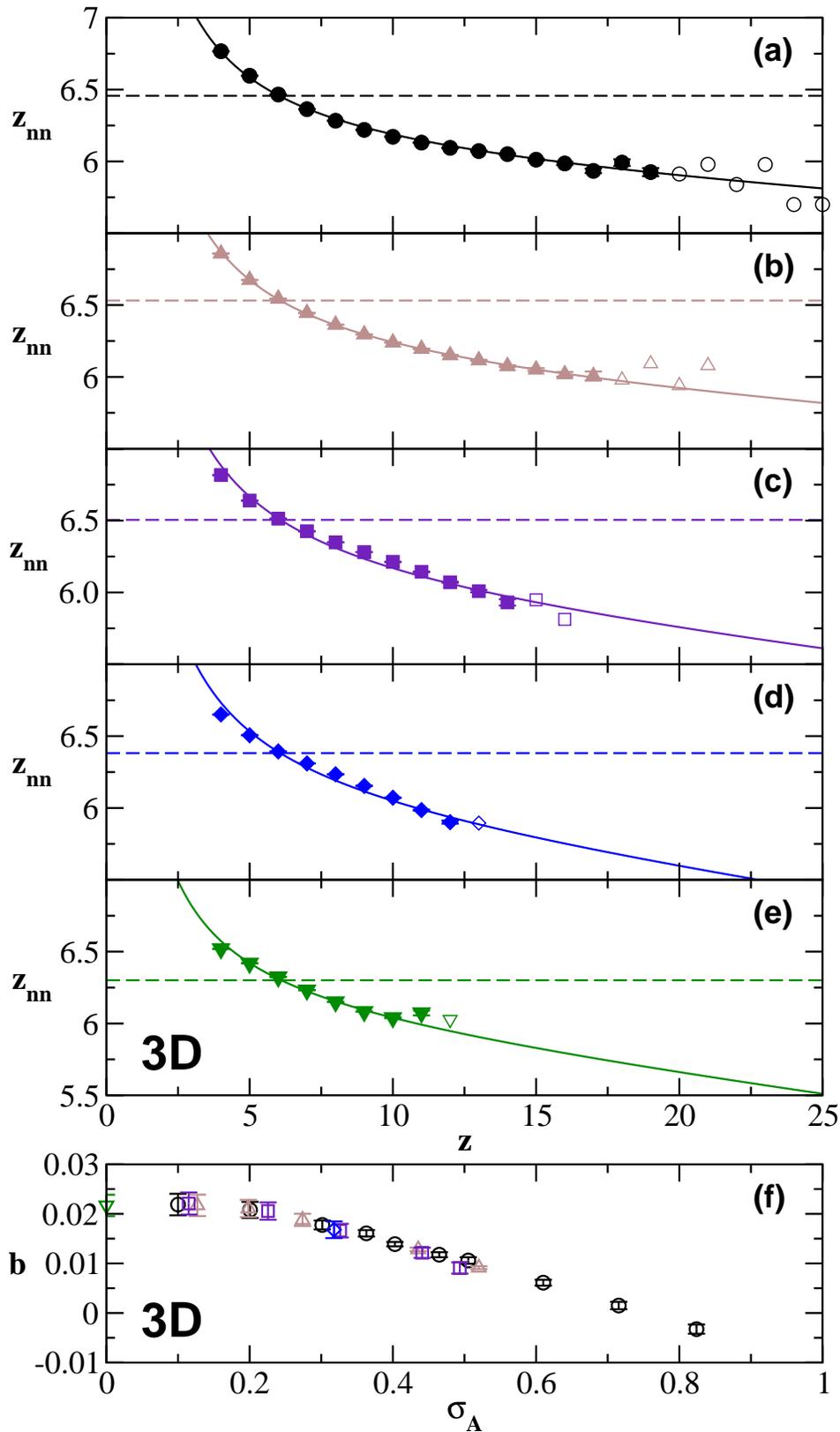}
\caption{\label{figAW5} Contact number correlations for spheres in contact at $\phi_{RCP}$. The error bars are standard deviations from the mean. The solid line are fits of Equation (\ref{eqonetrue}) to the data represented by closed symbols, open symbols are data omitted from fitting due to low occurrence. The dashed line is the prediction of an uncorrelated packing from Equation (\ref{eqZunc}). The data plotted in each panel is: (a) lognormal $\sigma_A = 0.40$; (b) Gaussian $\sigma_A = 0.44$; (c) uniform $\sigma_A = 0.44$; (d) bidisperse; (e) monodisperse. In (f) the fit parameter $b$ is plotted as a function $\sigma_A$. The data in (f) is labelled the same as in Figure \ref{figphiRCP}(a).}
\end{center}
\end{figure}

It is therefore worthwhile to investigate this in more detail. To do so we define $Z_{nn}(z)$, the average contact number of particles that are in contact with a particles that has $z$ contacts. $Z_{nn}(z)$ is analogous to the quantity $m(n)$, the average number of neighbours neighbouring a cell with $n$ neighbours, in the Aboav-Weaire law, which is well studied for two dimensional cellular structures\cite{Aboav,WeaireAboav}. Recently it has been shown that two dimensional disc packings exhibit nearest neighbour correlations in the contact network\cite{MobiusMe} similar to the anti-correlations observed in cellular structures.

In our previous work \cite{MobiusMe} we proposed an analogue of the Aboav-Weaire law for cellular structures to describe these correlation in two dimensional disc packings. This analogue draws upon a counting argument\cite{WeaireAboav} that leads to a sum rule that is exact and independent of dimension,
\begin{equation}
\label{Wsum}
\sum\limits_{z}(Z_{nn} z -z^2) P(z) = 0.
\end{equation}
$Z_{nn}$ is then a function of $z$ that must satisfy Equation (\ref{Wsum}). For uncorrelated packings $Z_{nn}(z)$ is a constant ($\equiv\overline{Z}_{nn}$) which is independent of $z$. Using Equation (\ref{Wsum}) we can show that it is given by 
\begin{equation}
\label{eqZunc}
\overline{Z}_{nn} = \avg{z} + \frac{\sigma_{Z}^{2}}{\avg{z}}.
\end{equation}
Figure \ref{figAW5} shows $Z_{nn}(z)$ for various size distributions. All distributions exhibit clear anti-correlations, namely that particles with few contacts are surrounded by particles with many contacts and vice versa. However, the deviations from the uncorrelated prediction $\overline{Z}_{nn}$ is usually less than $10\%$, which may explain the reason why the granocentric approach works well.

While there is currently no theoretical prediction for these correlations, one can find an empirical equation for $Z_{nn}(z)$ based on a series expansion of $Z_{nn}(z)$ in terms of the moments of $P(z)$. In order to ensure that the sum rule (Equation (\ref{Wsum})) is satisfied the series takes the form of
\begin{equation}
\label{classic}
(Z_{nn} - \avg{z})z -  \sigma_{Z}^{2} = -\sum_{i=1}c_i \left( z^i - \avg{z^i}\right),
\end{equation}
where the $c_i$'s are arbitrary constants. If $c_i$ = 0 for $i>1$, one recovers the Aboav-Weaire law for cellular structures. For 2D packings we found that the data was well described by only making the second term non-zero \cite{MobiusMe} which leads to a one parameter fit:
\begin{equation}
\label{eqonetrue}
Z_{nn} = \avg{z} - b z + \frac{b \avg{z}^2 + \sigma_{Z}^{2}(1+b)}{z},
\end{equation}
where $b = c_2$. This empirical equation agrees well with the correlations we find in 2D \cite{MobiusMe} and 3D packings as shown in Figure \ref{figAW5}.

The fit parameter $b$ does not depend on the shape of the size distribution but only on the width $\sigma_A$ as shown in Figure \ref{figAW5}(f). Note that all size distributions regardless of shape or width exhibit these anti-correlations in the contact network. 
 
\subsection{Nearest neighbour size correlations}

In the following we investigate spatial correlations between the size of particles in our polydisperse packings in two and three dimensions. Another way to look at this question is as follows: Is the size distribution of particles neighbouring a central particle of radius $r_c$ just equal to the global size distribution $P(r)$ as assumed in the granocentric approach?

\begin{figure}
\begin{center}
\includegraphics[width=120.0mm]{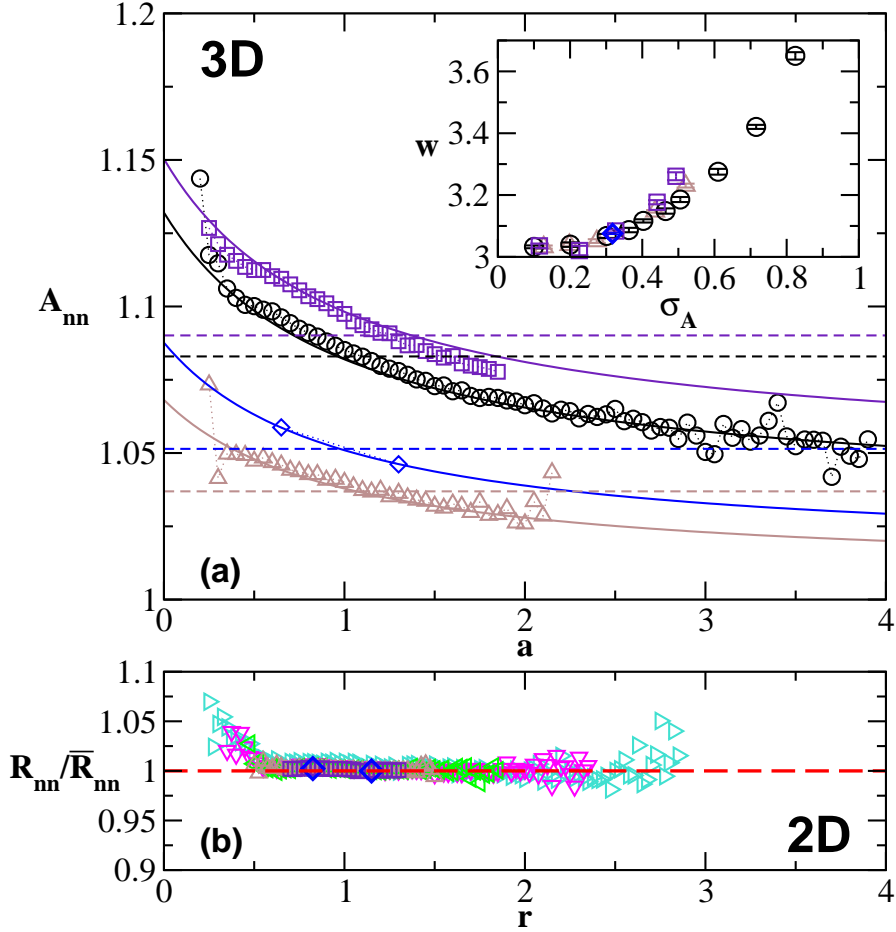}
\caption{\label{figAlaw}(a) Correlations between size of spheres in contact. All symbols represent the same size distributions as in Figure \ref{figAW5} except: (\textcolor{brown}{$\vartriangle$}) Gaussian $\sigma_A = 0.27$. The solid lines are fits to the data using Equation (\ref{eqAlaw}) with the dashed lines being the uncorrelated prediction calculated from Equation (\ref{eqauncz}). The inset shows the fit parameter $w$ as a function of $\sigma_A$. The data in the inset is labelled the same as in Figure \ref{figphiRCP}(a). (b) Correlations between size of discs in contact in two dimensions rescaled by the predicted uncorrelated radius $\overline{R}_{nn}$. Five different size distributions are plotted; (\textcolor{blue}{$\lozenge$}) bidisperse, radius ratio 1:1.4; (\textcolor{Purple}{$\square$}) uniform $\sigma_R = 0.17$; (\textcolor{brown}{$\vartriangle$}) Gaussian $\sigma_R = 0.14$; (\textcolor{green}{$\vartriangleleft$}) lognormal $\sigma_R = 0.20$; (\textcolor{magenta}{$\triangledown$}) lognormal $\sigma_R = 0.30$; (\textcolor{cyan}{$\triangleright$}) lognormal $\sigma_R = 0.40$}
\end{center}
\end{figure}

In order to explore potential size correlations, $A_{nn}(a)$ is defined as the average normalised surface area of all particles in contact with a particle with area $a$. Figure \ref{figAlaw}(a), shows $A_{nn}$ versus $a$ for four different size distributions. This result clearly indicates spatial anti-correlations in the particle size and are not consistent with the uncorrelated prediction $\overline{A}_{nn}$ which we discuss below. On average larger particles are surrounded by smaller particles and vice versa.

The same counting argument that is used to formulate Equation (\ref{Wsum}) can be applied to the particle size. Analogous to Equation (\ref{Wsum}), $A_{nn}$ must satisfy the following relation:
\begin{equation}
\label{Asum}
    \int_{0}^{\infty} A_{nn}(a) \avg{z|a} P(a) da = \int_{0}^{\infty} a \avg{z|a} P(a) da.
\end{equation}
Using a series expansion in terms of the moments of the area distribution $P(a)$ that satisfies Equation (\ref{Asum}) we obtain
\begin{equation}
\label{eqAnngen}
    A_{nn}(a)= \frac{\int_{0}^{\infty} a \avg{z|a} P(a) da + \sum_i w_{i} (a^{i} - \avg{a^{i}})}{\avg{z|a}},
  \end{equation}
  where $w_i$ are arbitrary constants. Previously we have shown that $\avg{z|a}$ was well described by a linear relation (Equation (\ref{eqzafit})). Substituting this expression into Equation (\ref{eqAnngen}) and keeping only the first term in the expansion $(i=1)$ yields,
    \begin{equation}
  \label{eqAlaw}
   A_{nn}(a)=\frac{\avg{z} + \gamma \sigma_{A}^{2} + w (a-1) }{\avg{z} + \gamma (a -1)},
  \end{equation}
  where $w = w_1$. This is a one parameter fit to the data, since the value of $\gamma$ is independent of polydispersity. The fits shown in Figure \ref{figAlaw}(a) agree well with the data for all the size distributions we considered. The inset shows the corresponding values of $w$ which mostly depend on the width but not the shape of the size distribution. Therefore, Equation (\ref{eqAlaw}) provides a good description of the correlations between the size of particles in 3D disordered packings.

\begin{figure}
\begin{center}
\includegraphics[angle=270,width=120.0mm]{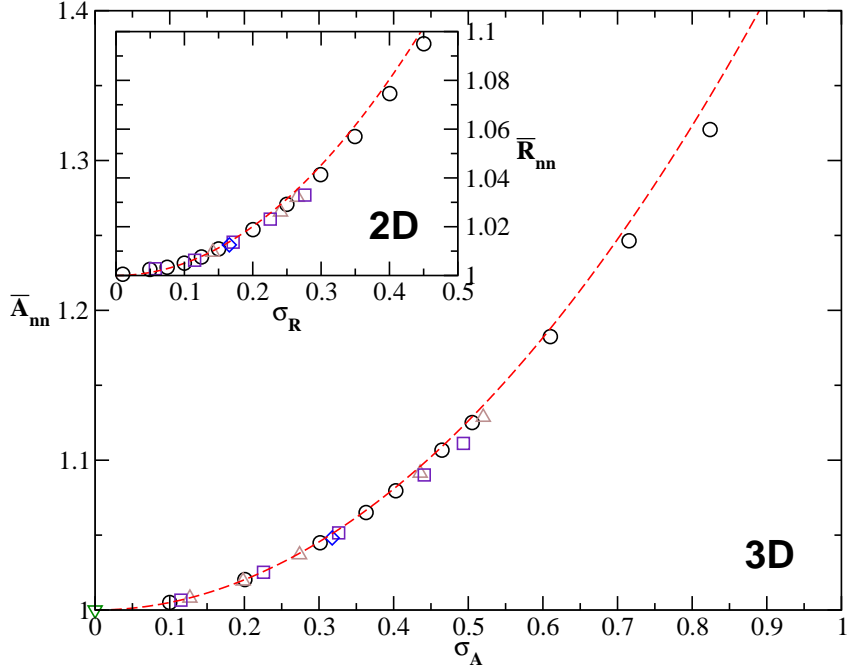}
\caption{\label{figaunc}The uncorrelated prediction $\overline{A}_{nn}$ for a range of different size distributions in three dimensions. The dashed line is the uncorrelated prediction of $\overline{A}_{nn}$ calculated from Equation (\ref{eqauncpred}). Inset: The uncorrelated prediction $\overline{R}_{nn}$ for a range of different size distribution in two dimensions. The dashed line is the uncorrelated prediction of $\overline{R}_{nn}$ calculated from Equation (\ref{eqruncpred}). The data is labelled the same as in Figure \ref{figphiRCP}(a).}
\end{center}
\end{figure}

 In the absence of correlations $A_{nn}(a)$ becomes a constant ($\equiv \overline{A}_{nn}$) which we can derive from Equation (\ref{Asum}):
 \begin{equation}
 \label{ann1}
 \overline{A}_{nn} = \int_{0}^{\infty} a \left(\frac{\avg{z|a}}{\avg{z}} P(a)\right) da.
 \end{equation}
 Substituting  Equation (\ref{eqzafit}) into the previous expression, it can be simplified to
 \begin{equation}
 \label{eqauncpred}
 \overline{A}_{nn} = 1 + \frac{\gamma}{\avg{z}} \sigma_{A}^{2},
 \end{equation}
 where $\gamma$ and $\avg{z}$ are constant at $\phi_{RCP}$. Therefore, $\overline{A}_{nn}$ depends on $\sigma_A$ only. This relation is plotted in Figure \ref{figaunc} and agrees well with the data. Wider distributions give rise to a larger $\overline{A}_{nn}$. 
 
Alternatively, the sum rule analogue for particle size can be written as the sum
 \begin{equation}
    \sum_z \avg{A_{nn}|z} z P(z)=\sum_z z \avg{a|z} P(z),
  \end{equation}
  where $\avg{A_{nn}|z}$ is the average area of all particles in contact with a particle that has $z$ contacts. In the absence of correlations $\avg{A_{nn}|z}$ reduces to 
\begin{equation}
\label{eqauncz}
    \overline{A}_{nn}=\frac{1}{\avg{z}}\sum_z z \avg{a|z} P(z).
  \end{equation}
  This expression for $\overline{A}_{nn}$ is equivalent to Equation (\ref{ann1}) as can be seen from Equation (\ref{eqazeq}).

We previously reported \cite{MobiusMe} a sum rule for nearest neighbour size correlations in two dimensions
\begin{equation}
\label{Rsum}
    \int_{0}^{\infty} R_{nn}(r) \avg{z|r} P(r) dr = \int_{0}^{\infty} r \avg{z|r} P(r) dr.
\end{equation}
where $R_{nn}(r)$ is the average normalised radius in contact with a particle of radius $r$. The data is shown in Figure \ref{figAlaw}(b). In contrast to the results for three dimensional sphere packings, the size distribution appears to be spatially uncorrelated. $R_{nn}$ is well described by
\begin{equation}
\label{eqruncpred0}
\overline{R}_{nn} = \int_{0}^{\infty} r \left(\frac{\avg{z|r}}{\avg{z}} P(r)\right) dr,
\end{equation}
for all polydipsersities. Using the the linear fit $\avg{z|r}$ this reduces to
\begin{equation}
\label{eqruncpred}
 \overline{R}_{nn} = 1 + \frac{\gamma_{2D}}{\avg{z}} \sigma_{R}^{2},
\end{equation}
where $\gamma_{2D}$ and $\avg{z}$ are constant at $\phi_{RCP}$.

Figure \ref{figaunc} shows that both $\overline{A}_{nn}$ and $\overline{R}_{nn}$ increase with $\sigma_A$ and $\sigma_R$, respectively. The data is well described by Equations (\ref{eqauncpred}) and  (\ref{eqruncpred}). The reason for both $\overline{A}_{nn}$ and $\overline{R}_{nn}$ to be greater than $1$ is the fact larger particles have more contacts and therefore appear more frequently in the first neighbour shell. 

These results affect the central assumption in the granocentric approach, namely that the size distribution of the neighbour shell is equivalent to the global $P(r)$. Even in the absence of correlations, as for 2D packings, the average particle radius of the first neighbour shell is larger than $1$ - the mean radius of the global size distribution. When correlations are neglected, the effective size distribution $P(r_{nn})$ of particles in the (contacting) neighbour shell is simply
\begin{equation}
P(r_{nn}) = \left[\frac{\avg{z|r}}{\avg{z}}P(r)\right]_{r=r_{nn}} ,
\end{equation}
which follows from Equation (\ref{eqruncpred0}). Consequently, the assumption that the size distribution of the neighbouring particles is governed by $P(r)$ becomes progressively worse for broader distribution.

In practice, the difference between the weighted size distribution shown above and $P(r)$ is small for moderate polydispersities and therefore does not measurably affect the outcome of the predictions made by the granocentric model. 

\section*{Conclusions}

We have shown that a surprising number of features of frictionless packings are insensitive to polydispersity. Our key result is the universal correlations we observe between size and contact number of a particle that are independent of the shape and width of the size distribution. This holds in both two and three dimensions and allows a mean field formulation of the granocentric model. The contact number distributions emerging from the model agree well with the data for a wide range of polydispersities. The two parameters that appear in the model do not depend on polydispersity either. 

In passing we note that the random close packing density excluding the rattlers remains unchanged for a wide range of discrete and continuous size distributions in both dimensions. This holds even for packings that contain up to $30\%$ of rattlers. 

Despite the success of the granocentric approach, we note that a central assumption in this model does not hold. We find that all packings exhibit anti-correlations in the contact network. In addition, the particle sizes are anti-correlated, but only in 3D packings. For three dimensional packings we can therefore conclude that on average smaller particles which have less contacts are surrounded by larger particles that have more contacts.

Nevertheless, the granocentric model, while not exact, yields good predictions since these correlations are typically weak.

\section*{Acknowledgements}
C. B. O'D would like to acknowledge funding from the School of Physics, Trinity College Dublin. The authors are grateful for stimulating discussions with S. Hutzler, D. Weaire and M. Clusel.

\bibliographystyle{tPHM}
\bibliography{foam_bib}

\end{document}